# PERMUTATION TESTS FOR EQUALITY OF DISTRIBUTIONS OF FUNCTIONAL DATA

by

Federico A. Bugni
Department of Economics
Duke University
Durham, NC 27708 USA

and

Joel L. Horowitz
Department of Economics
Northwestern University
Evanston, IL 60208 USA

June 2021

**Abstract**

Economic data are often generated by stochastic processes that take place in continuous time, though observations may occur only at discrete times. For example, electricity and gas consumption take place in continuous time. Data generated by a continuous time stochastic process are called functional data. This paper is concerned with comparing two or more stochastic processes that generate functional data. The data may be produced by a randomized experiment in which there are multiple treatments. The paper presents a method for testing the hypothesis that the same stochastic process generates all the functional data. The test described here applies to both functional data and multiple treatments. It is implemented as a combination of two permutation tests. This ensures that in finite samples, the true and nominal probabilities with which each test rejects a correct null hypothesis are equal. The paper presents upper and lower bounds on the asymptotic power of the test under alternative hypotheses. The results of Monte Carlo experiments and an application to an experiment on billing and pricing of natural gas illustrate the usefulness of the test.

---

We thank the co-editor and three anonymous referees for comments and suggestions that greatly improved this paper. Part of this research was carried out while Joel L. Horowitz was a visitor at the Department of Economics, University College London, and the Centre for Microdata Methods and Practice. The research of Federico Bugni was supported in part by NIH Grant 40-4153-00-0-85-399 and NSF Grant SES-1729280.

**PERMUTATION TESTS FOR EQUALITY OF DISTRIBUTIONS OF FUNCTIONAL DATA**

## 1. INTRODUCTION

Economic data are often generated by stochastic processes that take place in continuous time. Examples are gas and electricity consumption by households, asset prices or returns, and wages. Other continuous time processes include the temperature and concentration of an air pollutant at a given location. Many of these processes can be recorded in continuous time, though they may be reported only in discrete time. Data generated from a continuous time stochastic process are random functions and are called functional data. The analysis of functional data is a well-established research area in statistics that has generated a vast literature. See, for example, Hall and Hossein-Nasab (2006); Jank and Shmueli (2006); Ramsay and Silverman (2002, 2005); Yao, Müller, and Wang (2005); and the references therein.

In this paper, we are concerned with comparing two or more stochastic processes that generate functional data. These processes are produced by a randomized experiment in which there are one or more treatment groups and one control group. Our objective is to test the hypothesis that the same stochastic process generates the functional data in all the groups. More precisely, the null hypothesis is that the functional data (random functions) generated by the stochastic processes for the treatment and control groups have the same probability distribution. Our interest in this hypothesis is motivated by experiments in billing and pricing of gas and pricing of electricity that have been conducted in several countries, including the US and Ireland. In a typical experiment, households are assigned randomly to treatment and control groups. The treatment groups have one or more experimental billing or price schedules, and the control group has regular billing and pricing. Consumption of gas or electricity by households in the treatment and control groups is measured at frequent time intervals for several months. For example, in the Irish experiment on gas billing and pricing that we analyze later in this paper, consumption was measured every 30 minutes for twelve months. Gas consumption takes place in continuous time, though it is measured only at discrete times. The consumption path of a household is a random function of continuous time. The consumption paths of all households in the treatment groups (control group) are random samples of functions generated by the treatment (control) consumption processes. The hypothesis tested in this paper is that the consumption processes of the treatment and control groups are the same. The alternative hypothesis is that the treatment and control processes differ on a set of time intervals with non-zero Lebesgue measure.

If the hypothesis to be tested pertained to the distributions of finite-dimensional random variables, then testing could be carried out using the Cramér-von Mises or Kolmogorov-Smirnov two-sample tests, among others (Schilling 1986, Henze 1988), or multi-sample generalizations of these tests. But the Cramér-von Mises and Kolmogorov-Smirnov tests do not apply to random functions, which are infinite-dimensional random variables. Methods are also available for testing the hypothesis that continuous time data or, equivalently, random functions are generated by a known stochastic process or a process that is known up to a finite-dimensional parameter (Bugni, Hall, Horowitz, and Neumann 2009; Cuesta-Albertos, del Barrio, Fraiman, and Matrán 2007; Cuesta-Albertos, Fraiman and Ransford 2006; Hall and Tajvidi 2002; Kim and Wang 2006). Methods of parametric time-series analysis can also be used in this setting. However, the method described here is nonparametric. It does not assume that the stochastic processes generating the data have known parametric or semiparametric forms. Nor does it assume that the processes have a specific dependence structure such as stationarity or the error components structure of a panel. The dependence structure is completely general and unknown.

Another possibility is to carry out nonparametric tests of hypotheses of equality of specific features (e.g., moments) of the processes generated by the various treatment groups. For example, Harding and Lamarche (2016) compared moments of the distributions of electricity consumption in the treatment and control groups in a time-of-day pricing experiment. However, a test of equality of specific moments does not reveal whether the processes generated by the various groups differ in other ways. The method described in this paper facilitates such an investigation.

There are several existing methods for carrying out non-parametric distributional tests. Székely and Rizzo (2004, hereinafter SR) describe a test for data that may be high-dimensional but not functional. The test of SR is consistent, but its other asymptotic power properties are unknown. Schilling (1986) and Henze (1988) describe two-sample nearest neighbor tests for multivariate (not functional) data. The asymptotic power properties of these tests are unknown. However, we have found through Monte Carlo experiments that the power of several distributional tests increases as the frequency of discrete observations increases or, equivalently, the lengths of the intervals between discrete observations decrease. This motivates the development of tests that apply to functional data as well as discrete data. Hall and Tajvidi (2002, hereinafter HT) describe a permutation test for functional data. In principle, the HT test is an alternative to the test developed in this paper. However, the HT statistic depends on a user-chosen tuning parameter $\gamma$ and user-chosen weights $w$ that HT define. The HT statistic is highly sensitive to the choices of $\gamma$ and $w$, but there is no known systematic method for choosing these quantities in applications. In the empirical application described in Section 6 and the Monte Carlo experiments described in Section 7, we found that the $p$-value and power of the HT test can vary by factors of more than 10 and 4, respectively. This has led us to conclude that the HT test is not reliable in the settings of interest in this paper. Consequently, we do not use the HT test in the remainder of the paper.

The test described here is applicable to experiments with multiple treatment groups and a control group as well as experiments with one treatment group and a control group. This is an important property of the test. Experiments with multiple treatments are common in many fields (see, for example, Chong, Cohen, Field, Nakasone, and Torero (2016); Ashraf, Field, and Lee (2014); and Field, Jayachandran, Pande, and Rigol (2016), among many others). The experiment on gas billing and pricing analyzed later in this paper has multiple treatments.

The test statistic described here combines two tests. One of the tests is motivated by the statistic of Bugni, Hall, Horowitz, and Neumann (2009) (hereinafter BHHN), who describe a Cramér-von Mises-type test of the hypothesis that a sample of random functions was generated by a continuous time stochastic process that is known up to a finite-dimensional parameter. The first of the two tests used in the present paper is a Cramér-von Mises type test of the hypothesis that two or more samples of random functions were generated by the same unknown stochastic process. The alternative hypothesis is that the samples were generated by different stochastic processes.

The results of Monte Carlo experiments show that tests based on the Cramér-von Mises type statistic have relatively high power against alternative hypotheses consisting of changes in the variance and covariance functions of the stochastic processes under consideration but relatively low power against changes in the means of these processes. Therefore, we combine the Cramér-von Mises type test with a simple statistic for comparing the means of the stochastic processes. The results of Monte Carlo experiments show that our proposed combination of tests has higher power in important settings than a test based on the SR statistic.

The test is implemented as a combination of two permutation tests, which ensures that in a finite sample, the true and nominal probabilities with which each test rejects a correct null hypothesis are equal. A test based on the bootstrap or asymptotic approximations to the distribution of the test statistic does not have this property. The test proposed here has non-trivial power against alternative hypotheses that differ from the null hypothesis by $O(N^{-1/2})$, where $N$ is the number of observations in the largest sample. "Non-trivial" means that the power of the test exceeds the probability with which the test rejects a correct null hypothesis. The asymptotic local powers of the permutation tests are the same as they would be if the critical values of the tests were based on the asymptotic distributions of the test statistics under the null hypothesis. Thus, there is no penalty in terms of asymptotic power for the permutation test's elimination of the finite-sample error in the probability of rejecting a correct null hypothesis.

Section 2 of this paper presents the proposed test statistic for the case of a single treatment group and a control group. Section 2 explains how the critical values are obtained and describes the procedure for implementing the test. Section 3 presents the properties of two-sample version of the test under the null and alternative hypotheses. Section 4 extends the results of Sections 2 and 3 to experiments in which there are several treatment groups and a control group. Section 5 discusses methods for selecting a user-chosen measure that is used in the test. Section 6 applies the test to data from a multiple-treatment experiment on the pricing of gas. Section 7 presents the results of simulation studies of the test's behavior using a design that mimics the experiment analyzed in Section 6. Section 7 also presents simulation results illustrating the sensitivity of the HT test to the choice of tuning parameters. Section 8 presents concluding comments. The proofs of theorems are in the appendix. Section A.4 of the appendix presents the results of Monte Carlo experiments that illustrate the sensitivity of the HT test to user-chosen tuning parameters. Section A.4 also presents the results of experiments that illustrate the increase in the powers of the SR test and our test as the frequency of discrete observations increases.

## 2. THE NULL HYPOTHESIS AND TEST STATISTICS IN THE SINGLE TREATMENT CASE

### 2.1 The Test Statistic

Let $\mathcal{I} = [0, T]$ be a closed interval, and let $L_2(\mathcal{I})$ denote the set of real-valued, square-integrable functions on $\mathcal{I}$. In contrast to the usual definition of $L_2(\mathcal{I})$, we define two square-integrable functions that differ on a set of Lebesgue measure zero to be distinct. We consider two stochastic processes (or random functions) on $\mathcal{I}$: $X(t) \in L_2(\mathcal{I})$ and $Y(t) \in L_2(\mathcal{I})$. For example, $X(t)$ may correspond to the treatment group and $Y(t)$ to the control group. In the gas pricing experiment, $\mathcal{I}$ is the period of time over which gas consumption is observed. $X(t)$ and $Y(t)$, respectively, are gas consumption at time $t$ by individuals in the treatment and control groups. Let $F_X$ and $F_Y$ respectively be the probability distribution functions of $X(t)$ and $Y(t)$. That is, for any non-stochastic function $z$ that is square-integrable on $\mathcal{I}$,

(2.1) $$F_X(z) = P[X(t) \le z(t) \text{ for all } t \in \mathcal{I}]$$

and

(2.2) $$F_Y(z) = P[Y(t) \le z(t) \text{ for all } t \in \mathcal{I}].$$

The null hypothesis to be tested is

(2.3) $$H_0: F_X(z) = F_Y(z)$$

for all $z \in L_2(\mathcal{I})$. The alternative hypothesis we consider is

(2.4) $$H_1: P_\mu[F_X(Z) \ne F_Y(Z)] > 0,$$

where $\mu$ is a probability measure on $L_2(\mathcal{I})$ and $Z$ is a random function with probability distribution $\mu$. $H_1$ is equivalent to the hypothesis that $F_X(z) \ne F_Y(z)$ on a set of $z$'s with non-zero $\mu$ measure. The measure $\mu$ is analogous to a weight function in tests of the Cramer-von Mises type, among others. Like the weight function in other tests, $\mu$ in the test presented here influences the directions of departure from $H_0$ in which the test has high power. The choice of $\mu$ is discussed in Section 5.

Now define

$$\tau = \int [F_X(z) - F_Y(z)]^2 \, d\mu(z),$$

and define

$$\nu = \int_0^T [EX(t) - EY(t)]^2 \, dt$$

if the expectations exist. Then $\tau = \nu = 0$ under $H_0$, $\tau > 0$ under $H_1$, and $\nu > 0$ if $EX(t) \neq EY(t)$ on a set of non-zero Lebesgue measure. A Cramér-von Mises type test of $H_0$ can be based on a sample analog of $\tau$ that is scaled to have a non-degenerate limiting distribution. A test of $EX(t) \neq EY(t)$ can be based on a sample analog of $\nu$. The Cramér-von Mises type test is consistent against $H_1$. That is, the probability that the test rejects $H_0$ when $H_1$ is true approaches 1 as the sample size increases. However, as we discuss later in this paper, the test has low finite-sample power against mean shifts. In a mean shift, $X(t) - EX(t) = Y(t) - EY(t)$ for all $t$, but $EX(t) \neq EY(t)$. Therefore, we combine sample analogs of $\tau$ and $\nu$ to obtain our final test. Section 2.3 describes the combined test.[1]

To obtain the sample analog of $\tau$, let $\{X_i(t): i = 1,...,n\}$ and $\{Y_i(t): i = 1,...,m\}$ denote random samples (sample paths) of $n$ and $m$ realizations of $X(t)$ and $Y(t)$, respectively. Make

<u>Assumption 1</u>: (i) $X(t)$ and $Y(t)$ are separable, $\mu$-measurable stochastic processes. (ii) $\{X_i(t): i = 1,...,n\}$ is an independent random sample of the process $X(t)$. $\{Y_i(t): i = 1,...,m\}$ is an independent random sample of the process $Y(t)$ and is independent of $\{X_i(t): i = 1,...,n\}$.

Also assume for the moment that $X_i(t)$ and $Y_i(t)$ are observed for all $t \in \mathcal{I}$. The more realistic setting in which $X_i(t)$ and $Y_i(t)$ are observed only at a discrete set of points $t \in \mathcal{I}$ is treated in the next paragraph.[2] Define the empirical distribution functions

$$(2.5) \qquad \hat{F}_X(z) = n^{-1} \sum_{i=1}^{n} I[X_i(t) \leq z(t) \text{ for all } t \in \mathcal{I}]$$

---

[1] The power of Cramér von Mises type tests against mean shifts is explored through Monte Carlo experiments in Section 7.

[2] $X(t)$ and $Y(t)$ are stochastic processes, such as gas consumption, that take place in continuous time but can be observed (measured) only at discrete time points, say $t_1, t_2, ..., t_J$. A test of a hypothesis about the discrete-time processes $X(t_j)$ and $Y(t_j)$ ($j = 1,...,J$) is an approximation to a test of about the continuous time processes $X(t)$ and $Y(t)$ ($t \in \mathcal{I}$). The power of a test of a hypothesis about the discrete-time processes increases as the number of time points $J$ increases. See Table A1 in the appendix. Therefore, we develop a test that has desirable properties in the continuous time setting but can be used with discrete time.

and

(2.6) $$\hat{F}_Y(z) = m^{-1}\sum_{i=1}^{m} I[Y_i(t) \le z(t) \text{ for all } t \in \mathcal{I}] .$$

The sample analog of $\tau$ is

(2.7) $$\tau_{nm} = (n+m)\int [\hat{F}_X(z) - \hat{F}_Y(z)]^2 d\mu(z) .$$

$H_0$ is rejected if $\tau_{nm}$ is larger than can be explained by random sampling error. The integral in (2.7) may not have a closed analytic form. In that case, $\tau_{mn}$ can be replaced with a simulation estimator that is obtained by randomly sampling $\mu$. Let $\{Z_\ell : \ell = 1,...,L\}$ be such a sample. Then the simulation version of $\tau_{nm}$ is

(2.8) $$\hat{\tau}_{nm} = (n+m)L^{-1}\sum_{\ell=1}^{L}[\hat{F}_X(Z_\ell) - \hat{F}_Y(Z_\ell)]^2 .$$

Arguments like those used to prove Theorem 3.3 of BHHN can be used to show that $\hat{\tau}_{nm} \to^{a.s.} \tau_{nm}$ with respect to the probability measure $\mu$ as $L \to \infty$. However, the $\alpha$-level permutation test based on $\hat{\tau}_{nm}$ rejects a correct $H_0$ with probability exactly $\alpha$, even if $L$ is finite. See Theorem 3.1.

Now suppose that $X_i(t)$ and $Y_i(t)$ are observed only at the discrete times $\{t_j : j = 1,...,J;\ 0 \le t_j \le T\}$. Then the empirical distribution functions $\hat{F}_X$ and $\hat{F}_Y$ are replaced by

$$\tilde{F}_X[z(t_1),...,z(t_J)] = n^{-1}\sum_{i=1}^{n} I[X_i(t_j) \le z(t_j) \text{ for all } j = 1,...,J]$$

and

$$\tilde{F}_Y[z(t_1),...,z(t_J)] = m^{-1}\sum_{i=1}^{m} I[Y_i(t_j) \le z(t_j) \text{ for all } j = 1,...,J].$$ [3]

The test statistic remains as in (2.7), except the arguments of the empirical distribution functions are the finite-dimensional vector $[z(t_1),....,z(t_J)]'$. The test statistic is

$$\tau_{nm} = (n+m)\int \{\tilde{F}_X[z(t_1),...,z(t_J)] - \tilde{F}_Y[z(t_1),...,z(t_J)]\}^2 d\mu(z) .$$

Define $\zeta_j = z(t_j)$ ($j = 1,...,J$). Then $\tau_{nm}$ is equivalent to

(2.9) $$\tau_{nm} = (n+m)\int [\tilde{F}_X(\zeta_1,...,\zeta_J) - \tilde{F}_Y(\zeta_1,...,\zeta_J)]^2 f_J(\zeta_1,...,\zeta_J) d\zeta_1 ... d\zeta_J ,$$

---

[3] $\tilde{F}_X$ and $\tilde{F}_Y$ produce the version of our test that results from choosing a discrete measure. The resulting test is identical to the test in (2.7) with a discrete measure.

where $f_J$ is the probability density function on $\mathbb{R}^J$ induced by $\mu$.

We now present the sample analog of $\nu$. Make

Assumption 2: $EX(t)$ and $EY(t)$ exist and are finite for all $t \in [0,T]$.

If $X_i(t)$ and $Y_i(t)$ are observed for all $t \in \mathcal{I}$, define

$$\hat{E}X(t) = n^{-1}\sum_{i=1}^{n} X_i(t)$$

and

$$\hat{E}Y(t) = m^{-1}\sum_{i=1}^{n} Y_i(t).$$

The sample analog of $\nu$ is

$$\nu_{nm} = (n+m)\int_0^T [\hat{E}X(t) - \hat{E}Y(t)]^2 dt. \quad (2.10)$$

If $X_i(t)$ and $Y_i(t)$ are observed only at the discrete times $\{t_j : j = 1,...,J;\ 0 \le t_j \le T\}$, then

$$\nu_{nm} = (n+m)J^{-1}\sum_{j=1}^{J}[\hat{E}X(t_j) - \hat{E}Y(t_j)]^2. \quad (2.11)$$

## 2.2 Critical Values and the Test Procedure

Under $H_0$ and mild regularity conditions, the empirical processes $(n+m)^{1/2}[\hat{F}_X(z) - \hat{F}_Y(z)]$ and $(n+m)^{1/2}[\hat{E}X(t) - \hat{E}Y(t)]$ converge weakly to mean-zero Gaussian processes. In addition, $(n+m)^{1/2}[\tilde{F}_X(\zeta_1,...,\zeta_J) - \tilde{F}_Y(\zeta_1,...,\zeta_J)]$ and $(n+m)^{1/2}\{[\hat{E}X(t_1) - \hat{E}Y(t_1)],...,[\hat{E}X(t_J) - \hat{E}Y(t_J)]\}'$ are asymptotically normal. These results can be used to derive the asymptotic distributions of $\tau_{nm}$ and $\nu_{nm}$ under $H_0$ with either continuous-time or discrete-time observations of $X(t)$ and $Y(t)$. The asymptotic distributions can be used in the usual way to obtain asymptotic critical values of $\tau_{nm}$ and $\nu_{nm}$. However, asymptotic approximations can be inaccurate and misleading in finite samples. We avoid this problem by carrying out permutation tests based on $\tau_{nm}$ and $\nu_{nm}$. Lehmann and Romano (2015, Ch. 15) provide a general discussion of such tests. The critical value of a permutation test does not depend on asymptotic approximations. The true and nominal probabilities of rejecting a correct null hypothesis with a permutation test are equal in finite samples. Moreover, the asymptotic power of the permutation test is the same as the power the test based on the asymptotic critical value. This section explains the permutation test procedure and how to obtain critical values for permutation tests based on $\tau_{nm}$. As is explained in Section 2.1, the same results apply to the simulation version of $\tau_{mn}$. Critical values for permutation tests based on $\nu_{nm}$ can be obtained by replacing $\tau_{nm}$ with $\nu_{nm}$ throughout the following discussion.

Let $\alpha \in (0,1)$ be the nominal level of the $\tau_{nm}$ test. The $\alpha$-level critical value is computed by evaluating $\tau_{nm}$ for permutations of the combined sample of $n+m$ observations of $\{X_i :\ i=1,...,n;\ Y_i :\ i=1,...,m\}$. There are $Q=(m+n)!$ ways of dividing the $(n+m)$ observations in the combined sample into one set of $m$ observations and another of $n$ observations.[4] Let $q=1,...,Q$ index these divisions or permutations, and let $\tau_{nmq}$ denote the test statistic based on the $q$'th permutation. The $\alpha$-level critical value of $\tau_{mn}$ is the $(1-\alpha)$ quantile of $\tau_{nmq}$ over $q=1,...,Q$. Denote this by $t^*_{nm}(1-\alpha)$. Then,

$$t^*_{nm}(1-\alpha) = \inf\left\{ t \in \mathbb{R} :\ Q^{-1}\sum_{q=1}^{Q} I(\tau_{nmq} \le t) \ge 1-\alpha \right\}.$$

If $Q$ is large, then $t^*_{nm}(1-\alpha)$ can be estimated with arbitrary accuracy by replacing the sums over all $Q$ permutations of the observations with sums over a random sample of $\tilde{Q}$ permutations. The $\alpha$-level test rejects a correct $H_0$ with probability exactly $\alpha$, even if $t^*_{nm}(1-\alpha)$ is estimated by this random sampling method (Lehmann and Romano 2005, p. 636).

Among the $(n+m)!$ permutations of the data, only the $(n+m)!/(n!m!)$ combinations consisting of one group of $n$ observations and another of $m$ observations yield distinct values of $\tau_{nmq}$. Therefore, the permutation test can be defined in terms of combinations of the data, rather than permutations. The critical value and properties of the test are the same, regardless of whether $\tau_{nmq}$ is defined using permutations or combinations.

To carry out the permutation test based on $\tau_{nm}$, define

$$\varphi_{nm} = \begin{cases} 1 \text{ if } \tau_{nm} > t^*_{nm}(1-\alpha) \\ a \text{ if } \tau_{nm} = t^*_{nm}(1-\alpha) \\ 0 \text{ if } \tau_{nm} < t^*_{nm}(1-\alpha) \end{cases}$$

where

[4] The ordering of the observations in a permutation does not matter. Therefore, the same test is obtained if one replaces $Q$ with the $\binom{m+n}{m}$ distinct combinations of the $(n+m)$ observations.

$$a = \frac{Q\alpha - Q^+}{Q^0},$$

$$Q^+ = \sum_{q=1}^{Q} I[\tau_{nmq} > t_{nm}^*(1-\alpha)],$$

and

$$Q^0 = \sum_{q=1}^{Q} I[\tau_{nmq} = t_{nm}^*(1-\alpha)].$$

Let $\phi_{nm}$ denote a Bernoulli random variable that is equal to one with probability $\varphi_{nm}$. By construction, a test that rejects $H_0$ if $\phi_{nm} = 1$ is a size $\alpha$ test. Such a test has a random outcome when $\tau_{nm} = t_{nm}^*(1-\alpha)$. A possibly conservative non-stochastic level $\alpha$ test can be obtained by replacing $a$ above with 0.

2.3 The Combined Test

To form the combined test, let $\alpha_\tau, \alpha_\nu \in (0,1)$. Let $t_{nm}^*(1-\alpha_\tau)$ and $\phi_{nm}$ be the quantities defined in Section 2.2 but with $\alpha_\tau$ in place of $\alpha$. Let $\nu_{nm}^*(1-\alpha_\nu)$ and $\tilde{\phi}_{nm}$ be the same quantities but with $\nu_{nm}$ in place of $\tau_{nm}$ and $\alpha_\nu$ in place of $\alpha$. The combined test rejects $H_0$ if $\eta_{nm} \equiv \max(\phi_{nm}, \tilde{\phi}_{nm}) > 0$. Thus, the combined test rejects $H_0$ if either the $\tau_{nm}$ test or the $\nu_{nm}$ test rejects $H_0$. The probability that the combined test rejects $H_0$ is $P[(\phi_{nm} > 0) \cup (\tilde{\phi}_{nm} > 0)]$. It follows from the Bonferroni inequality that under $H_0$

$$\max(\alpha_\tau, \alpha_\nu) \le P[(\phi_{nm} > 0) \cup (\tilde{\phi}_{nm} > 0)] \le \alpha_\tau + \alpha_\nu. \quad (2.12)$$

Thus, for example, if $\alpha_\tau = 0.04$ and $\alpha_\nu = 0.01$, the combined test rejects a correct $H_0$ with probability between 0.04 and 0.05. Section 7 presents the results of Monte Carlo experiments that illustrate the power of the combined test with several combinations of $\alpha_\tau$ and $\alpha_\nu$ satisfying $\alpha_\tau + \alpha_\nu = 0.05$. The results of the experiments, which are shown in Table 3, indicate that the level and rejection probability of the combined test are not very sensitive to the choices of $\alpha_\tau$ and $\alpha_\nu$.[5]

---

[5] An alternative approach is to base a permutation test on a single statistic that combines $\tau_{nm}$ and $\nu_{nm}$. However, $\tau_{nm}$ and $\nu_{nm}$ have different scales. Combining them in a way that does not cause one statistic to dominate the other requires putting them on a common scale, possibly through Studentization or prepivoting. In the context of permutation-based tests, Chung and Romano (2013), Fogarty (2020), and Bertanha and Chung (2021) discuss Studentization. Chung and Romano (2016) and Fogarty (2021) discuss prepivoting in the same context. We have found that Studentization of our statistics is analytically complicated and computationally burdensome, especially with functional data or many observation points. Prepivoting eliminates analytical complexity at the expense of significantly increasing the computational cost. In contrast, the permutation test proposed in Section 2.3 is much

## 3. PROPERTIES OF THE TEST IN THE SINGLE TREATMENT CASE

### 3.1 Finite Sample Properties under $H_0$

The following theorem gives the finite-sample behavior of $\tau_{nm}$ and $\nu_{nm}$ under $H_0$ with the critical values $t_{nm}^*(1-\alpha_\tau)$, and $\tilde{t}_{nm}^*(1-\alpha_\nu)$, respectively.

Theorem 3.1: Let assumptions 1 and 2 hold. For any distribution $P$ that satisfies $H_0$ and any $\alpha_\tau, \alpha_\nu \in (0,1)$,

$$E_P(\phi_{nm}) = \alpha_\tau ,$$

and

$$E_P(\tilde{\phi}_{nm}) = \alpha_\nu . \blacksquare$$

Theorem 3.1 implies that the true and nominal rejection probabilities of the tests based on $\tau_{nm}$ and $\nu_{nm}$ are equal regardless of:

1. The measure $\mu$ or probability density function $f_J$ that is used to define $\tau_{nm}$.
2. Whether $X_i(t)$ and $Y_i(t)$ are observed in continuous time or only at discrete points in time.
3. Whether the integrals in (2.7) and (2.9) are calculated in closed form or estimated by simulation as in (2.8).
4. Whether $t_{nm}^*$ and $\tilde{t}_{nm}^*$ are computed using all $Q$ possible permutations of the data or only an independent random sample of $\tilde{Q} < Q$ permutations.

Theorem 3.1 also implies that $\eta_{nm}$ satisfies (2.12) when $H_0$ is true.

### 3.2 Asymptotic Properties when $H_0$ is False

This section presents conditions under which tests based on $\tau_{nm}$, $\nu_{nm}$, and $\eta_{nm}$ reject a false $H_0$ with probability approaching 1 as $n,m \to \infty$. The asymptotic local power functions of tests based on $\tau_{nm}$ and $\nu_{nm}$ are presented in the appendix. These functions show that under assumption 3 below, the $\tau_{nm}$ test has non-trivial power against alternatives whose distance from the null hypothesis is $O(n^{-1/2})$. The $\nu_{nm}$ test has non-trivial power against alternatives for which the distance measure $\int E[X(t)-Y(t)]^2 dt$ is $O(n^{-1/2})$. "Non-trivial power" means that the probability of rejecting a false null hypothesis exceeds the probability of rejecting a correct one. It follows from the definition of $\eta_{nm}$ that the asymptotic local and

---

easier to compute and implement, though at the cost of being slightly conservative. We consider this to be an acceptable price to pay for a procedure that is computationally convenient and, therefore, potentially appealing to applied researchers.

finite sample powers of tests based on $\eta_{nm}$ equal or exceed the powers of separate tests based on $\tau_{nm}$ and $\nu_{nm}$ at the $\alpha_\tau$ and $\alpha_\nu$ levels, respectively. The asymptotic local power functions of $\tau_{nm}$ and $\nu_{nm}$ are very complicated and, consequently, not useful for comparing the local powers of $\tau_{nm}$ and $\nu_{nm}$ with each other and with the local powers of other tests. Section 7 presents the results of a Monte Carlo investigation of the powers of the tests.

To obtain the limiting probabilities with which tests based on $\tau_{nm}$ and $\nu_{nm}$ reject a false $H_0$, make

Assumption 3: As $n \to \infty$, $m = m(n) \to \infty$ and $m / n \to \lambda$ for some finite $\lambda > 0$.

The following theorems give conditions for consistency of the $\tau_{nm}$ and $\nu_{nm}$ tests against a false $H_0$ when $X(t)$ and $Y(t)$ are observed in continuous time or at discrete times.

Theorem 3.2: Let assumptions 1 and 3 hold.

a. If $X(y)$ and $Y(t)$ are observed in continuous time, $0 < \alpha_\nu < 1$, and

$$\int [F_X(z) - F_Y(z)]^2 d\mu(z) > 0, \tag{3.1}$$

then

$$\lim_{n\to\infty} P[\tau_{nm} > t_{nm}^*(1-\alpha_\tau)] = 1.$$

b. If $X(t)$ and $Y(t)$ are observed at the discrete time points $t_1,...,t_J$, $0 < \alpha_\nu < 1$, and $\mu$ concentrates on points $\{z(t_j): j = 1,...,J\}$, then the conclusion of part a. holds. ■

Theorem 3.3: Let assumptions 1-3 hold.

a. If $X(y)$ and $Y(t)$ are observed in continuous time, $0 < \alpha_\nu < 1$, and

$$\int_0^T [EX(t) - EY(t)]^2 dt > 0,$$

then

$$\lim_{n\to\infty} P[\nu_{nm} > \tilde{t}_{nm}^*(1-\alpha_\nu)] = 1.$$

b. If $X(y)$ and $Y(t)$ are observed at the discrete time points $t_1,...,t_J$, $0 < \alpha_\nu < 1$, and

$$\sum_{j=1}^{J} [EX(t_j) - EY(t_j)]^2 > 0,$$

then

$$\lim_{n\to\infty} P[\nu_{nm} > \tilde{t}_{nm}^*(1-\alpha_\nu)] = 1. \quad ■$$

Theorem 3.2 implies that the test based on $\eta_{nm}$ rejects $H_0$ with probability approaching 1 as $n \to \infty$ if condition (3.1) holds.

## 4. EXTENSION TO MULTIPLE TREATMENTS

This section outlines the extension of the results of Sections 2 and 3 to the case in which there are two or more treatment groups and a single control group. We assume that the outcomes of all treatment groups are continuously observed. Results for $\tau_{nm}$ and discretely observed outcomes can be obtained by replacing the measure $\mu$ for continuously observed outcomes with a measure that concentrates on the observed times $\{t_j: j = 1,...,J\}$. Results for $\nu_{nm}$ and discretely observed outcomes can be obtained by replacing (2.10) with (2.11). As in Section 2.3, the test based on the multiple treatment extension of $\eta_{nm}$ rejects $H_0$ if the multiple-treatment extension of either $\tau_{nm}$ or $\nu_{nm}$ rejects $H_0$.

Let $s = 0,1,...,S$ index treatment groups with the control group labelled $s = 0$. Let $X_s(t)$ denote the outcome process in treatment group $s$. For each $s = 0,...,S$ define the cumulative distribution function

$$F_s(z) = P[X_s(t) \le z(t) \text{ for all } t \in \mathcal{I}].$$

The null hypothesis is

$$H_0: \ F_s = F_0 \ \text{ for all } s = 1,...,S.$$

The alternative hypothesis is

$$H_1: \ P_\mu[F_s(Z) \ne F_0(Z) \text{ for some } s = 1,...,S] > 0.$$

Let $\{X_{is}(t): i = 1,...,n_s\}$ denote a random sample (sample paths) of $n_s$ realizations of $X_s(t)$. Define $n = \sum_{s=0}^{S} n_s$. The following assumptions extend assumptions 1-3 to the case of multiple treatments.

<u>Assumption 1′</u>: (i) $X_s(t)$ ($s = 0,...,S$) is a separable, $\mu$-measurable stochastic process. (ii) For each $s = 0,..,S$, $\{X_{is}(t): i = 1,...,n_s\}$ is an independent random sample of the process $X_s(t)$. Moreover, $X_{is}(t)$ and $X_{j\tilde{s}}(t)$ are independent of each other if $(i,s) \ne (j,\tilde{s})$.

<u>Assumption 2′</u>: $EX_s(t)$ exists and is finite for all $t \in [0,T]$ and $s = 0,...,S$.

<u>Assumption 3′</u>: For each $s$ there is a constant $\pi_s > 0$ such that $n_s / n \to \pi_s$ as $n \to \infty$.

For each $s = 0,...,S$ define the empirical distribution function

$$\hat{F}_s(z) = n_s^{-1} \sum_{i=1}^{n} I[X_{is}(t) \le z(t) \text{ for all } t \in \mathcal{I}].$$

Let $\mu$ be the measure defined in Section 2.1, and define $\boldsymbol{n} = (n_0, n_1, ..., n_S)'$. The extensions of $\tau_{nm}$ and $\nu_{nm}$ to the multiple treatment case are

$$\tau_{\boldsymbol{n}} = \sum_{s=1}^{S} (n_0 + n_s) \int [\hat{F}_0(z) - \hat{F}_s(z)]^2 d\mu(z)$$

and

$$\nu_{\boldsymbol{n}} = \sum_{s=1}^{S} (n_0 + n_s) \int_0^T [\hat{E}X_0(t) - \hat{E}X_s(t)]^2 dt .$$

The multiple-treatment test is implemented by permuting the observed sample paths so that there are $n_s$ permuted observations in treatment group $s$. Let $\tau_{\boldsymbol{n}q}$ and $\nu_{\boldsymbol{n}q}$ denote the statistics obtained from permutation $q$. The critical values of $\tau_{\boldsymbol{n}}$ and $\nu_{\boldsymbol{n}}$ are obtained using the method described in Section 2.2 with $\tau_{nmq}$ and $\nu_{nmq}$, respectively, replaced by $\tau_{\boldsymbol{n}q}$ and $\nu_{\boldsymbol{n}q}$. Denote the $\alpha$-level critical values by $t_{\boldsymbol{n}}^*(1-\alpha)$ and $\tilde{t}_{\boldsymbol{n}}^*(1-\alpha)$. As in the single-treatment case, the $\alpha$-level multiple-treatment tests based on $\tau_{\boldsymbol{n}}$ and $\nu_{\boldsymbol{n}}$ reject a correct $H_0$ with probability exactly $\alpha$. Let $\eta_{\boldsymbol{n}} \equiv \max(\phi_n, \tilde{\phi}_n) > 0$ if the combined test rejects $H_0$ and $\eta_{\boldsymbol{n}} = 0$ otherwise. The combined test rejects a correct $H_0$ with probability between $\max(\alpha_\tau, \alpha_\nu)$ and $\alpha_\tau + \alpha_\nu$.

The multiple-treatment analogs of the continuous time versions of Theorems 3.2 and 3.3 are:

<u>Theorem 4.1</u>: Let assumptions 1′ and 3′ hold. If $X_s(t)$ is observed in continuous time, $0 < \alpha < 1$, and

$$\int \sum_{s=1}^{S} [F_s(z) - F_0(z)]^2 d\mu(z) > 0 , \tag{4.1}$$

then

$$\lim_{n \to \infty} P[\tau_{\boldsymbol{n}} > t_{\boldsymbol{n}}^*(1-\alpha)] = 1 . \ \blacksquare$$

<u>Theorem 4.2</u>: Let assumptions 1′-3′ hold. If $X_s(t)$ is observed in continuous time for all $s$, $0 < \alpha < 1$, and

$$\sum_{s=1}^{S} \int_0^T [EX_0(t) - EX_s(t)]^2 dt > 0 ,$$

then

$$\lim_{n \to \infty} P[\nu_{\boldsymbol{n}} > \tilde{t}_{\boldsymbol{n}}^*(1-\alpha_\nu)] = 1 . \ \blacksquare$$

Theorem 4.1 implies that the multiple treatment extension of the combined test rejects $H_0$ with probability approaching 1 as $n \to \infty$ if condition (4.1) holds. Similar results apply to the discrete-time versions of the multiple treatment tests.

## 5. THE MEASURE $\mu$

As was stated in Section 2.1, the measure $\mu$ influences the directions of departure from $H_0$ in which tests based on $\tau_{nm}$ and $\tau_{\boldsymbol{n}}$ have high power. This section presents informal suggestions about how $\mu$ can be constructed. We emphasize that regardless of the choice of $\mu$, the probability that $\alpha$-level permutation tests based on $\tau_{nm}$ and $\tau_{\boldsymbol{n}}$ reject a correct null hypothesis is exactly $\alpha$. A more formal approach to constructing $\mu$ is outlined at the end of this section.

To obtain a flexible class of measures, let $\{\psi_k : k = 1, 2, ...\}$ be a complete, orthonormal basis for $L_2[\mathcal{I}]$. We use a basis of trigonometric functions in Sections 6 and 7, but the same measure can be generated with any complete, orthonormal basis. Let $\mu$ be the probability measure generated by the random function

$$Z(t) = \sum_{k=1}^{\infty} b_k \psi_k(t), \tag{5.1}$$

where the Fourier coefficients $\{b_k : k = 1, 2, ...\}$ are random variables satisfying

$$\sum_{k=1}^{\infty} b_k^2 < \infty \tag{5.2}$$

with probability 1. Sample paths $Z_j(t)$ are generated randomly by sampling the $b_k$'s randomly from some distribution such that (5.2) holds with probability 1. The distribution of the $b_k$'s implies the measure $\mu$. Therefore, $\mu$ can be specified by specifying the basis functions $\{\psi_k\}$ and the distribution of the Fourier coefficients $\{b_k\}$, which ensures that $\mu$ is a probability distribution on $L_2(\mathcal{I})$. The test statistic can be computed using (2.8) by truncating the infinite sum in (5.1) at some integer $K$, randomly sampling the $b_k$'s and computing $Z_i(t)$'s as

$$Z_i(t) = \sum_{k=1}^{K} b_{ki} \psi_k(t),$$

where $b_{ki}$ is the $i$'th realization of the random variable $b_k$.

The mean of $Z(t)$ is

$$E[Z(t)] = \sum_{k=1}^{K} E(b_k) \psi_k(t).$$

An investigator who expects $|F_X[z(t)] - F_Y[z(t)]|$ to be relatively large in certain ranges of $t$ can choose $E[Z(t)]$ to be a function, say $w(t)$, that is large in those ranges and set

$$E(b_k) = \int_0^T w(t)\psi_k(t)dt .$$

An investigator who has no such expectations might choose $w(t)$ to be a constant. Given a choice of $w(t)$ and the resulting mean Fourier coefficients $E(b_k)$, the $b_k$ 's can be specified as

$$b_k = E(b_k) + \rho_k U_k ,$$

where the $U_k$ 's are random variables that are independently and identically distributed across values of $k$ with $E(U_k) = 0$ and $Var(U_k) = 1$, and the $\rho_k$ 's are non-stochastic constants satisfying

$$\sum_{k=1}^{\infty} \rho_k^2 < \infty .$$

The distributions of the $U_k$ 's can set equal to $U[-3^{1/3}, 3^{1/3}]$ or $N(0,1)$ if the distributions of the processes $X(t)$ and $Y(t)$ have thin tails. If $X(t)$ and $Y(t)$ have heavy-tailed distributions, then one might consider taking the variables $U_k$ to have heavy-tailed distributions such as Student-$t$ with a low number of degrees of freedom.

A more formal approach to choosing $\mu$ is to specify an alternative hypothesis, specify the distributions of the Fourier coefficients $b_k$ up to finitely many parameters, and choose the parameters through Monte Carlo simulation to maximize power. The computation required to implement this approach is difficult and time-consuming, because the objective function of the optimization problem is non-convex and must be evaluated through high-dimensional numerical integration. We carried out the power-optimization approach with several of the Monte Carlo designs described in Section 7 and found that it produced little increase in power over the informal choice of $\mu$ described in Section 6.

## 6. AN EMPIRICAL APPLICATION

This section reports the application of the $\eta_n$ test to data produced by the smart metering consumer behavior trial (CBT) for gas conducted by the Commission for Energy Regulation (CER) of Ireland. The CER is Ireland's independent regulator of electricity and natural gas. The goal of the CBT was to investigate the effects of several different billing and pricing treatments on residential customers' consumption of gas. The gas consumption of each customer in the CBT was measured every half hour by a smart meter. The CER kindly provided the data produced by the CBT and related documentation (Commission for Energy Regulation 2011).

The CBT was divided into two periods, a baseline period that took place from December 2009 through May 2010 and an experimental period that took place from June 2010 through May 2011. During the baseline period, all customers participating in the CBT were charged the standard rate for gas and were billed bimonthly in the usual way. During the experimental period, customers were assigned randomly to a control group or one of four treatment groups. Customers then received different treatments depending on their assignments. Customers in the control group continued to be charged the standard rate and billed bimonthly. Customers in the first treatment group were charged at the standard rate and billed bimonthly but also received a detailed report on their energy usage with recommendations about how to reduce consumption. Customers in the second treatment group were charged the standard rate but billed monthly instead of bimonthly. Customers in the third treatment group were charged at the standard rate and billed bimonthly but also received an in-home electronic device that displayed their instantaneous gas consumption and its cost. Customers in the fourth treatment group, like those in the third group, were billed bimonthly and received the in-home device. In addition, these customers were charged a variable rate according to the seasonal wholesale cost of procuring gas. Depending on the season, the rate these customers were charged was between 16 percent below the standard rate (in June through September 2010) and 17 percent above the standard rate (in December 2010 and January 2011).

The analysis in this section is concerned with gas consumption during the experimental period, when customers received different treatments depending on their assignment. We test the null hypothesis that the distributions of gas consumption by customers in the four treatment groups and the control group were the same in each month from June through December 2010. The data consist of observations of the gas consumption of 1492 customers at half-hour intervals. The numbers of customers in the treatment and control groups are shown in Table 1.

Figures 1-3 provide an informal illustration of the differences between the distributions of gas consumption in the five groups. Figure 1 shows average monthly gas consumption by customers in the control and four treatment groups; Figure 2 shows the average standard deviation of customers' consumption; and Figure 3 shows the average correlation coefficient of consumption in consecutive half-hour periods. It can be seen that the differences among the means and standard deviations of consumption in the different treatment groups are small, but there are larger differences among the correlation coefficients. Thus, the main effect of the experimental treatments appears to be a shift in the dependence structure of gas consumption.

We applied the multiple group version of the $\eta_n$ test and the test of SR to consumption in each of the months from June through December. The statistic $\eta_n$ combines $\tau_n$ and $\nu_n$. We used a trigonometric basis in (5.1) and a truncated series expansion to compute $\tau_n$. Thus, (5.1) became

$$Z(t) = b_1 + \sum_{k=1}^{(K-1)/2} \sqrt{2} b_{2k} \cos[k\pi(2t-T)/T] + \sum_{k=1}^{(K-1)/2} \sqrt{2} b_{2k+1} \sin[k\pi(2t-T)/T],$$

where $K$ is an integer and $T$ is the number of half hours in a month. The Fourier coefficients were

$$b_1 \sim N\left(\mu_1, 1/K\right),$$

where

$$\mu_1 = \text{median}_i \max_t \{X_i(t):\ i = 1,...,1492;\ t = 1,...,T\}$$

and

$$b_k \sim N\left(0, 1/K\right);\ k > 1\,.$$

The parameter $\mu_1$ is the mean of $Z(t)$ and is set near the center of the support of the data. Our test would have low power if $\mu_1$ were outside of or too close to the boundaries of the support. We computed $p$-values for the $\eta_n$ test for $K = 3,5,...,25$ and found little variation over this range. Therefore, we report only $p$-values for $K = 25$. The integrals in the definition of $\tau_n$ are population averages of functionals of $Z(t)$. We used $L = 4000$ draws of $Z(t)$ to approximate these integrals. Equation (2.8) shows the approximation for the single-treatment case. The approximation for multiple treatments, as in the CBT, is similar. We used 500 permutations of the data to compute critical and $p$-values for the $\tau_n$, $\eta_n$, and SR tests. We used four different $(\alpha_\tau, \alpha_V)$ pairs to compute $\eta_n$: $(0.02, 0.03)$, $(0.025, 0.025)$, $(0.03, 0.02)$, and $(0.04, 0.01)$. These values yield the $0.05$ significance level. To obtain any other significance level $\alpha$, we rescaled the foregoing values so as to achieve the desired $\alpha$ without changing $\alpha_V / \alpha_\tau$. For example, $(\alpha_\tau, \alpha_V) = (0.02, 0.03)$ was rescaled to $(\alpha_V, \alpha_\tau) = (0.02, 0.03)\alpha / 0.05$. The corresponding $p$-value is the value of $\alpha$ at which $\eta_n$ switches from 0 to 1.

The results of the tests are shown in Table 2. Rows 1-4 of Table 2 show the $p$-values obtained with the $\eta_n$ test. Row 5 shows the $p$-values obtained with the $\tau_n$ test. Row 6 shows the $p$-values obtained using the test of SR. All $p$-values are in percentages. The $\eta_n$ test rejects the null hypothesis of no treatment effect at the $0.05$ level in August. It rejects the null hypothesis at the $0.10$ level in July when $(\alpha_\tau, \alpha_V) = (0.04, 0.01)$ and $(\alpha_\tau, \alpha_V) = (0.03, 0.02)$ but not for the other values of $(\alpha_\tau, \alpha_V)$. The $\tau_n$ test rejects the null hypothesis of no treatment effect at the 0.05 level in August and at the 0.10 level in July and September. The $\eta_n$ test does not reject the null hypothesis in September through December. The SR test does not reject the null hypothesis in any of the months June through December ($p > 0.118$ in each month). The $\eta_n$, $\tau_n$, and SR tests are permutation tests, so all have correct finite-sample sizes. Therefore, the results shown in Table 2 indicate that the $\eta_n$ and $\tau_n$ tests detect a treatment effect that is not detected by the SR test.

## 7. MONTE CARLO EXPERIMENTS

This section reports the results of Monte Carlo experiments that explore the finite-sample properties of the $\tau_n$ and $\eta_n$ tests. The designs of the experiments are based on the empirical illustration of Section 6. We simulate observations of half-hour gas consumption during a 30-day month. Thus, $\mathcal{I} = \{1,...,T\}$ with $T = 1440$ half hours. Each simulated dataset consists of $n = 150$ individuals who are distributed evenly among a control group and two treatment groups. Thus, $s = 0,1,2$, $n_0 = n_1 = n_2 = 50$, and $n = \sum_{s=0}^{2} n_s = 150$. Each simulated dataset $\{X_{is}(t): t \in \mathcal{I};\ i = 1,...,n_s;\ s = 0,1,2\}$ was generated as follows:

1. Draw random variables $\{\xi_{is}(t): (i,t) \in \{1,...,n\} \times \mathcal{I};\ s = 0,1,2\}$ independently from the $N(0,1)$ distribution.
2. For all $i = 1,...,n_s$ and $s = 0,1,2$; set $\tilde{X}_{is}(0) = \xi_{is}(0)$.
3. For all $i = 1,...,n_s$; $s = 0,1,2$; and $t \in \mathcal{I}$, set $\tilde{X}_{is}(t) = \rho_s(t)\tilde{X}_{is}(t-1) + \xi_{is}(t)\sqrt{1-\rho_s^2(t)}$, where $\rho_s(t)$ is a parameter defined below.
4. For all $i = 1,...,n_s$; $s = 0,1,2$; and $t \in \mathcal{I}$, set $X_{is}(y) = \mu_s(t) + \sigma_s(t)\tilde{X}_{is}(t)$, where $\mu_s(t)$ and $\sigma_s(t)$ are parameters defined below.

The resulting random variables $\{X_{is}(t): t \in \mathcal{I};\ i = 1,...,n_s;\ s = 0,1,2\}$ are normally distributed with

1. $E[X_{is}(t)] = \mu_s(t)$.
2. $Var[X_{is}(t)] = \sigma_s^2(t)$.
3. $Corr[X_{is}(t), X_{is}(t-1)] = \rho_s(t)$ for all $t \in \mathcal{I}$ with $t > 1$.

In addition, $X_{i_1 s_1}(t_1)$ is independent of $X_{i_2 s_2}(t_2)$ if $i_1 \neq i_2$ or $s_1 \neq s_2$.

The specification of the experimental design is completed by defining the parameters $\mu_s(t)$, $\sigma_s(t)$, and $\rho_s(t)$. We chose the parameters of the control group ($s = 0$) to correspond to the CBT data in June 2010. For values of $t$ corresponding the first half hour of the day ($t = 1, 49, 97,...$) we set $[\mu_0(t), \sigma_0(t), \rho_0(t)]$ equal to the averages of those parameters in the CBT data over the first half hours of days in June 2010. For values of $t$ corresponding to the second half hour of each day ($t = 2, 50, 98,...$) we set $[\mu_0(t), \sigma_0(t), \rho_0(t)]$ equal to the averages of those parameters in the CBT data over the second half hours of days in June 2010. The values of $[\mu_0(t), \sigma_0(t), \rho_0(t)]$ for the remaining half hours were set similarly. The values of $[\mu_s(t), \sigma_s(t), \rho_s(t)]$ $(s = 1,2)$ for the two treatment groups varied according to the experiment. We did experiments with 10 different sets of values of $[\mu_s(t), \sigma_s(t), \rho_s(t)]$, which we call parameter designs. The 10 parameter designs are:

1. No treatment effect: $[\mu_s(t), \sigma_s(t), \rho_s(t)] = [\mu_0(t), \sigma_0(t), \rho_0(t)]$ for all $t$ and $s = 1,2$.

2. Mean shift for treatment group 1: $[\mu_1(t), \sigma_1(t), \rho_1(t)] = [\mu_0(t) + 0.05, \sigma_0(t), \rho_0(t)]$ and $[\mu_2(t), \sigma_2(t), \rho_2(t)] = [\mu_0(t), \sigma_0(t), \rho_0(t)]$.

3. Mean shift for both treatment groups: $[\mu_1(t), \sigma_1(t), \rho_1(t)] = [\mu_2(t), \sigma_2(t), \rho_2(t)]$ $= [\mu_0(t) + 0.05, \sigma_0(t), \rho_0(t)]$.

4. Mean shift for treatment group 1 and variance shift for treatment group 2: $[\mu_1(t), \sigma_1(t), \rho_1(t)] = [\mu_0(t) + 0.05, \sigma_0(t), \rho_0(t)]$ and $[\mu_2(t), \sigma_2(t), \rho_2(t)] = [\mu_0(t), \sigma_0(t) + 0.05, \rho_0(t)]$.

5. Mean shift for treatment group 1 and correlation shift for treatment group 2: $[\mu_1(t), \sigma_1(t), \rho_1(t)] = [\mu_0(t) + 0.05, \sigma_0(t), \rho_0(t)]$ and $[\mu_2(t), \sigma_2(t), \rho_2(t)] = [\mu_0(t), \sigma_0(t), \rho_0(t) + 0.2]$.

6. Variance shift for treatment group 1: $[\mu_1(t), \sigma_1(t), \rho_1(t)] = [\mu_0(t), \sigma_0(t) + 0.05, \rho_0(t)]$ and $[\mu_2(t), \sigma_2(t), \rho_2(t)] = [\mu_0(t), \sigma_0(t), \rho_0(t)]$.

7. Variance shifts for both treatment groups: $[\mu_1(t), \sigma_1(t), \rho_1(t)] = [\mu_2(t), \sigma_2(t), \rho_2(t)]$ $= [\mu_0(t), \sigma_0(t) + 0.05, \rho_0(t)]$.

8. Correlation shift for treatment group 1: $[\mu_1(t), \sigma_1(t), \rho_1(t)] = [\mu_0(t), \sigma_0(t), \rho_0(t) + 0.2]$ and $[\mu_2(t), \sigma_2(t), \rho_2(t)] = [\mu_0(t), \sigma_0(t), \rho_0(t)]$.

9. Correlation shift for treatment group 1 and variance shift for treatment group 2: $[\mu_1(t), \sigma_1(t), \rho_1(t)] = [\mu_0(t), \sigma_0(t), \rho_0(t) + 0.2]$ and $[\mu_2(t), \sigma_2(t), \rho_2(t)] = [\mu_0(t), \sigma_0(t) + 0.05, \rho_0(t)]$.

10. Correlation shift for both treatment groups: $[\mu_1(t), \sigma_1(t), \rho_1(t)] = [\mu_2(t), \sigma_2(t), \rho_2(t)]$ $= [\mu_0(t), \sigma_0(t), \rho_0(t) + 0.2]$.

There were 1000 Monte Carlo replications in each experiment. Each experiment consists of computing the empirical probability that the null hypothesis of no treatment effect is rejected at the nominal 0.05 level. We compare the rejection probabilities of the $\tau_n$ and $\eta_n$ tests with the those of the SR test . We used several different values of $\alpha_\tau$ and $\alpha_\nu$ in the $\eta_n$ test. In all cases, $\alpha_\tau + \alpha_\nu = 0.05$ to ensure that the probability of rejecting a correct $H_0$ does not exceed 0.05. The power of the $\tau_n$ test and, therefore, of the $\eta_n$ test, depends on $K$. Accordingly, we carried out experiments with $K = 3,5,...,25$. In all experiments, the power of both tests is a monotonically increasing function of $K$ that becomes flat when $K$ reaches approximately 20. Therefore, we display powers only for $K = 25$. All experiments used $L = 4000$ in the tests based on $\tau_n$ and $\eta_n$.[6]

The results of the experiments are shown in Table 3. The results with design 1 indicate that the tests all have empirical probabilities of rejecting a correct null hypothesis that are close to the nominal probability. All of the tests are permutation tests, so this result is expected. The SR test is more powerful than the $\tau_n$ test in parameter designs 2-3, which consist of a mean shift. The $\eta_n$ and SR tests have similar powers in design 4, which has a variance shift in addition to a mean shift. In designs 5-10, which include variance or correlation shifts or both, the $\eta_n$ test is more powerful than the SR test. The SR test has especially low power in designs 6-10. These designs include variance and correlation shifts but not mean shifts. The SR test has high power only in the designs that include mean shifts.

The power of $\eta_n$ against mean shifts is similar to or higher than the power of SR when $\alpha_\tau = 0.025$ and $\alpha_\tau = 0.02$, whereas the $\tau_n$ test has low power against mean shifts. This motivates the test based on $\eta_n$, which combines $\tau_n$ with $\nu_n$. As is suggested by the low power of the $\tau_n$ test against mean shifts, the power of the $\eta_n$ against mean shifts is lower when $\alpha_\tau = 0.04$ than when $\alpha_\tau$ has lower values. The power of the $\eta_{nm}$ test against alternatives that do not involve mean shifts is highest when $\alpha_\tau = 0.04$. The results of the experiments show that the $\eta_n$ test overcomes the weakness of the $\tau_n$ test against mean shifts without substantially reducing power against variance and correlation shifts. We conclude that the test based on $\eta_n$ has good overall power compared to the SR test. The test based on $\eta_n$ is particularly good at detecting correlation shifts. We believe that this explains the empirical results of Section 6, as the CBT experimental treatment changed the correlation structure of gas consumption but had little effect on the mean or variance.

## 8. CONCLUSIONS

Economic data are often generated by stochastic processes that take place in continuous time, though observations may occur only at discrete times. Data generated by a continuous time stochastic process are called functional data. This paper has been concerned with comparing two or more stochastic processes that generate functional data. The data may be produced by a randomized experiment in which there are multiple treatments. The paper has presented a permutation test of the hypothesis that the same stochastic process generates all the functional data. The test described here applies to functional data and multiple treatments. The results of Monte Carlo experiments and an application to an experiment on billing and pricing of natural gas have illustrated the usefulness of the test.

[6] We also carried out experiments with permutation tests based on the nearest neighbor statistic of Schilling (1986) and Henze (1988). The powers of tests based on this statistic were much lower than the powers of tests based on the other statistics. Consequently, we do not show Monte Carlo results for the tests based on the nearest neighbor statistic.

## APPENDIX: PROOFS OF THEOREMS AND ADDITIONAL MONTE CARLO RESULTS

Section A.1 presents the proofs of Theorems 3.1-3.3, 4.1, and 4.2. Section A.2 presents auxiliary lemmas that are used in the proofs of the theorems. Section A.3 presents theorems giving the asymptotic local power functions of tests based on $\tau_{nm}$ and $\nu_{nm}$. Section A.4 presents the results of a Monte Carlo investigation of the sensitivity of the $\tau_n$, $\eta_n$, SR, and HT tests to various tuning parameters.

### A.1 Proofs of Theorems 3.1-3.3, 4.1, and 4.2

We present proofs only for $\tau_{nm}$. The proofs for $\nu_{nm}$ are the same after replacing $\tau_{nm}$ with $\nu_{nm}$.

Let $\boldsymbol{G}_{nm}$ denote the group of $Q=(m+n)!$ permutations of the $m+n$ observations $\{X_i: i=1,...,n;\ Y_i: i=1,...,m\}$ that produce one set of $n$ observations and another of $m$ observations. Let $(\mathcal{X}_n, \mathcal{Y}_m) = \{X_i: i=1,...,n;\ Y_i: i=1,...,m\}$ denote the original sample and $(\mathcal{X}_{nq}, \mathcal{Y}_{mq})$ denote the $q$'th permutation. Then

$$(\mathcal{X}_{nq}, \mathcal{Y}_{mq}) = g(\mathcal{X}_n, \mathcal{Y}_m)$$

for some function $g \in \boldsymbol{G}_{nm}$.

<u>Proof of Theorem 3.1</u>: For any $w \in \text{supp}(\mathcal{W}_{nm})$, the $\alpha$-level permutation test based on $\tau_{nm}$ can be written

$$\text{(A.1)} \qquad \varphi(w) = \begin{cases} 1 \text{ if } \hat{T}(w) > \hat{T}^{(k)}(w) \\ a(w) \text{ if } \hat{T}(w) = \hat{T}^{(k)}(w) \\ 0 \text{ if } \hat{T}(w) < \hat{T}^{(k)}(w), \end{cases}$$

where $\hat{T}(w)$ denotes $\tau_{nm}$ when $\mathcal{W}_{nm} = w$, $\hat{T}^{(k)}(w)$ denotes the $k$'th largest value of $\{\hat{T}(gw)\}_{g\in\boldsymbol{G}}$,

$$k = Q - \sup\{\gamma \in \mathbb{N}:\ \gamma \le Q\alpha\},$$

$$Q^0(w) = \sum_{g\in\boldsymbol{G}} I[\hat{T}(gw) = \hat{T}^{(k)}(w)],$$

$$Q^+(w) = \sum_{g\in\boldsymbol{G}} I[\hat{T}(gw) > \hat{T}^{(k)}(w)],$$

and

$$a(w) = [Q\alpha - Q^+(w)]/Q^0(w).$$

Let $\hat{T}^{(k)}(gw)$ denote the $k$'th largest value of $\hat{T}(gw)$. For each $g \in \boldsymbol{G}$, $\hat{T}^{(k)}(w) = \hat{T}^{(k)}(gw)$, $Q^0(gw) = Q^0(w)$, and $Q^+(gw) = Q^+(w)$. Consequently, $a(gw) = a(w)$. Moreover,

$$\varphi(gw) = \begin{cases} 1 \text{ if } \hat{T}(gw) > \hat{T}^{(k)}(w) \\ a(w) \text{ if } \hat{T}(gw) = \hat{T}^{(k)}(w) \\ 0 \text{ if } \hat{T}(gw) < \hat{T}^{(k)}(w) \end{cases}$$

and

$$\sum_{g \in \boldsymbol{G}} \varphi(gw) = Q^+(w) + a(w)Q^0(w) = Q^+(w) + \frac{Q\alpha - Q^+(w)}{Q^0(w)} Q^0(w) = Q\alpha .$$

Therefore, if $\mathcal{W}_{nm} \sim P$ for some distribution $P$ supported on $\text{supp}(\mathcal{W}_{nm})$, then

(A.2) $\quad Q^{-1} \sum_{g \in \boldsymbol{G}} E_P \{\varphi[g(\mathcal{W}_{nm})]\} = \alpha$.

Under $H_0$, $\mathcal{W}_{nm}$ is an independently and identically distributed sample of size $n+m$ with cumulative distribution function $F_X = F_Y = F$. Therefore, for any permutation $g \in G$, $\mathcal{W}_{nm} \sim g(\mathcal{W}_{nm})$. Consequently,

(A.3) $\quad E_P[\varphi(\mathcal{W}_{nm})] = E_P\{\varphi[g(\mathcal{W}_{nm})]\}$.

The theorem follows by combining (A.2), (A.3), and $Q = |\boldsymbol{G}|$. Q.E.D.

Proposition A.1: Define $\mathcal{W}_{nm} = (\mathcal{X}_n, \mathcal{Y}_m)$, and let $P_{nm}$ denote the probability distribution of $\mathcal{W}_{nm}$. Let $G_{nm}$ and $G'_{nm}$ be random variables that are uniformly distributed on $\boldsymbol{G}_{nm}$ independently of $\mathcal{W}_{nm}$ and each other. Let $\tau_{nm}(G_{nm}\mathcal{W}_{nm})$ denote the test statistic $\tau_{nm}$ evaluated using the transformed observations $G_{nm}\mathcal{W}_{nm}$. Suppose that under the sequence of probability measures $\{P_{nm} : n, m = 1,...,\infty\}$ and as $n, m \to \infty$,

$$[\tau_{nm}(G_{nm}\mathcal{W}_{nm}), \tau_{nm}(G'_{nm}\mathcal{W}_{nm})] \to^d (\tau, \tau'),$$

where $\tau$ and $\tau'$ are independently and identically distributed random variables with cumulative distribution function $R(\cdot)$. Define

$$r(1-\alpha) = \inf\{t \in \mathbb{R} : R(t) \geq 1-\alpha\}.$$

Then,

1. As $n, m \to \infty$, $\hat{R}_{nm}(t) \to^p R(t)$ for every $t$ that is a continuity point of $R$.
2. If $R(t)$ is continuous and strictly increasing at $t = r(1-\alpha)$, then

$$t^*_{nm}(1-\alpha) \to^p r(1-\alpha)$$

as $n, m \to \infty$.

3. Let $\tau_{nm} \to^d Z$ as $n,m \to \infty$, where $Z$ is a random variable with cumulative distribution function $\mathcal{J}$. Then

(a) $$\lim_{s \to r(1-\alpha)^-} \mathcal{J}(s) \le \liminf_{n,m\to\infty} P_{nm}[\tau_{nm} \le t^*_{nm}(1-\alpha)]$$

$$\le \limsup_{n,m\to\infty} P_{nm}[\tau_{nm} \le t^*_{nm}(1-\alpha)] \le \mathcal{J}[r(1-\alpha)]$$

(b) If $\mathcal{J}(t)$ is continuous at $t = r(1-\alpha)$, then

$$\lim_{n,m\to\infty} P_{nm}[\tau_{nm} \le t^*_{nm}(1-\alpha)] = \mathcal{J}[r(1-\alpha)]. \blacksquare$$

Proof: Parts 1 and 2 are proved by Lehmann and Romano (2005, Theorem 15.2.3). Part 3(a) is similar to Lemma 5 of Andrews and Guggenberger (2010). Part 3(b) is a corollary of part (a). Q.E.D.

The following notation is used in Lemma A.2, which is stated in the next paragraph. Let $\mathcal{O}$ denote the fixed subset of $\mathcal{I} = [0,T]$ on which $X(t)$ and $Y(t)$ are observed. $\mathcal{O} = \{t_1,...,t_J\}$ if $X(t)$ and $Y(t)$ are observed only at the discrete times $t_1,...,t_J$. $\mathcal{O} = [0,T]$ if $X(t)$ and $Y(t)$ are observed in continuous time. Let $F_X(z,\mathcal{O}) = P[X(t) \le z(t) \;\; \forall\, t \in \mathcal{O}]$ and $F_Y(z,\mathcal{O}) = P[Y(t) \le z(t) \;\; \forall\, t \in \mathcal{O}]$. For any function $D(z)$ satisfying $\int D(z)^2 d\mu < \infty$, define

$$F_{nX}(z,\mathcal{O}) = F_Y(z,\mathcal{O}) + (n+m)^{-1/2} D(z).$$

Let $\{\psi_k : k = 1,2,...\}$ be a complete orthonormal basis for $L_2(\mu)$ with the properties that are specified after (A.4) below. Let $\tilde{\Upsilon}(\zeta)$ be a Gaussian process indexed by $\zeta \in \mathbb{R}^J$ that has mean zero and covariance function

$$\text{cov}[\tilde{\Upsilon}(\zeta), \tilde{\Upsilon}(\tilde{\zeta})] = [(1+\lambda)^2/\lambda]\{F_Y[\min(\zeta,\tilde{\zeta})] - F_Y(\zeta)F_Y(\tilde{\zeta})\},$$

where $\min(\zeta,\tilde{\zeta})$ is the $J \times 1$ vector whose $j$'th component ($j = 1,...,J$) is $\min(\zeta_j,\tilde{\zeta}_j)$. Define $\Upsilon(z)$ as a Gaussian process indexed by $z \in L_2(\mu)$ with mean zero and covariance function

$$\text{cov}[\Upsilon(z), \Upsilon(\tilde{z})] = [(1+\lambda)^2/\lambda]\{F_Y[\min(z,\tilde{z})] - F_Y(z)F_Y(\tilde{z})\},$$

where $\min(z,\tilde{z})$ is the function of $t \in \mathcal{I}$ defined by $\min(z,\tilde{z}) = \{\min[z(t),\tilde{z}(t)] : t \in \mathcal{I}\}$. Let $\Upsilon^*$ be the process $\tilde{\Upsilon}$ defined if $\mathcal{O} = \{t_1,...,t_J\}$ and the process $\Upsilon$ if $\mathcal{O} = [0,T]$. Then

$$\int \Upsilon^*(z)\{\psi_k(z)\}_{k=1}^K d\mu(z) \sim N(\boldsymbol{0}_{K\times 1}, \Sigma_K)$$

for any positive integer $K$, where $\Sigma_K$ is the $K \times K$ matrix whose $(k,\tilde{k})$ component is (A.4) $(\Sigma_K)_{k,\tilde{k}} = [(\lambda+1)^2/\lambda]\int_z\int_{\tilde{z}}\{F_Y[\min(z,\tilde{z});\mathcal{O}] - F_Y(z;\mathcal{O})F_Y(\tilde{z};\mathcal{O})\}\psi_k(z)\psi_{\tilde{k}}(\tilde{z})d\mu(z)d\mu(\tilde{z})$.

The basis $\{\psi_k : k = 1,2,...\}$ can always be chosen so that

$$\sum_{k=1}^{\infty}(\Sigma_K)_{k,k} < \infty .$$

Define $N = n + m$, and let $\mathcal{W} = \{W_i : i = 1,...,N\}$ denote the combined samples of observations of $X$ and $Y$ with $W_i = X_i$ if $1 \le i \le n$ and $W_i = Y_i$ if $n+1 \le i \le n+m$.

Lemma A.2: Let assumptions 1 and 2 hold. Let $q_N$ and $\tilde{q}_N$ be two permutations of $\{1,...,N\}$ that are sample independently from the uniform distribution on $\{1,2,...,N\}$. Then

$$(\tau_{nmq_N}, \tilde{\tau}_{nmq_N}) \to^d (\tau, \tilde{\tau}),$$

where $\tau$ and $\tilde{\tau}$ are independently distributed as $\int \Upsilon^*(z)^2 d\mu(z)$. ■

Proof: For any permutation $q \in \{1,...,Q\}$ of $\{1,2,...,N\}$, let $i_q$ denote the position in the permutation of observation $i$ of $\mathcal{W}$. For any function $z(t)$ ($t \in \mathcal{O}$) define

$$H_{Nq}(z) = N^{-1/2}\sum_{i=1}^{N} U_{iq} I[W_i(t) \le z(t) \;\; \forall\, t \in \mathcal{O}],$$

where

$$U_{iq} = (N/n)I(i_q \le n) - (N/m)I(i_q > n).$$

Then

$$(\tau_{nmq_N}, \tau_{nm\tilde{q}_N}) = \left[\int H_{Nq_N}(z)^2 d\mu(z), \int H_{N\tilde{q}_N}(z)^2 d\mu(z)\right]$$

By the Cramér-Wold device, it suffices to show that

$$\alpha\tau_{Nq_N} + \beta\tau_{N\tilde{q}_N} \to^d \alpha\tau + \beta\tilde{\tau}.$$

for any constants $\alpha$ and $\beta$. For any positive integer $K$ and any $q \in \{q_N, \tilde{q}_N\}$,

$$H_{Nq}(z) = \sum_{k=1}^{\infty} c_{Nqk}\psi_k(z)$$

$$= H_{NqK1}(z) + H_{NqK2}(z),$$

where

$$c_{Nqk} = \int H_{Nq}(z)\psi_k(z)d\mu(z),$$

$$H_{NqK1}(z) = \sum_{k=1}^{K} c_{Nqk}\psi_k(z),$$

and

$$H_{NqK2}(z) = \sum_{k=K+1}^{\infty} c_{Nqk}\psi_k(z).$$

Also define

$$\tau_{NqK1} = \int H_{NqK1}(z)^2 d\mu(z) = \sum_{k=1}^{K} c_{Nqk}^2$$

and

$$\tau_{NqK2} = \int H_{NqK2}(z)^2 d\mu(z) = \sum_{k=K+1}^{\infty} c_{Nqk}^2 ,$$

where the second equality in the both lines follows from orthonormality of $\{\psi_k :\ k = 1,2,...\}$. Similarly,

(A.5) $$\Upsilon^*(z) = \sum_{k=1}^{\infty} b_k\psi_k(z) = \Upsilon^*_{K1}(z) + \Upsilon^*_{K2}(z),$$

where

(A.6) $$b_k = \int \Upsilon^*(z)\psi_k(z)d\mu(z),$$

(A.7) $$\Upsilon^*_{K1}(z) = \sum_{k=1}^{K} b_k\psi_k(z),$$

and

(A.8) $$\Upsilon^*_{K2}(z) = \sum_{k=K+1}^{\infty} b_k\psi_k(z).$$

Also define

(A.9) $$\tau_{K1} = \int \Upsilon^*_{K1}(z)^2 d\mu(z) = \sum_{k=1}^{K} b_k^2$$

and

(A.10) $$\tau_{K2} = \int \Upsilon^*_{K2}(z)^2 d\mu(z) = \sum_{k=K+1}^{\infty} b_k^2 .$$

Let $\tilde{\Upsilon}^*(z)$ be a process that is independent of but has the same distribution as $\Upsilon^*(z)$. Define $\tilde{\Upsilon}^*_{K1}(z)$, $\tilde{\Upsilon}^*_{K2}(z)$, $\tilde{b}_k$, $\tilde{\tau}_{K1}$, and $\tilde{\tau}_{K2}$ by replacing $\Upsilon^*(z)$ with $\tilde{\Upsilon}^*(z)$ in (A.5)-(A.10). To prove the theorem, it suffices to show that

(A.11) $\alpha\tau_{K1} + \beta\tilde{\tau}_{K1} \to^d \alpha\tau + \beta\tilde{\tau}$

as $K \to \infty$,

(A.12) $\alpha\tau_{Nq_N K1} + \beta\tau_{N\tilde{q}_N K1} \to^d \alpha\tau_{K1} + \beta\tilde{\tau}_{K1}$

as $N \to \infty$ for any positive integer $K$, and

(A.13) $\alpha(\tau_{Nq_N} - \tau_{Nq_N K1}) + \beta(\tau_{N\tilde{q}_N} - \tau_{N\tilde{q}_N K1}) = \alpha\tau_{Nq_N K2} + \beta\tau_{N\tilde{q}_N K2} \to^p 0$

as $N \to \infty$ followed by $K \to \infty$.

We begin with (A.11). It suffices to show that $\tau_{K1} \to^p \tau$ and $\tilde{\tau}_{K1} \to^p \tilde{\tau}$ as $K \to \infty$. We show that $\tau_{K1} \to^p \tau$. The same argument shows that $\tilde{\tau}_{K1} \to^p \tilde{\tau}$. Now $\tau - \tau_{K1} = \tau_{K2}$, so (A.7) follows from

$$E(\tau_{K2}) = \sum_{k=K+1}^{\infty} E(d_k^2) \to 0$$

as $K \to \infty$ because $E[\Upsilon^*(z)] \in L_2(\mu)$.

Next we show that (A.12) holds. For any positive integer $K$ define

$$C_{NK} = \left( \{c_{Nq_N k}\}_{k=1}^{K}, \{c_{N\tilde{q}_N k}\}_{k=1}^{K} \right)$$

and

$$B_K = \left( \{b_k\}_{k=1}^{K}, \{\tilde{b}_k\}_{k=1}^{K} \right).$$

Let $\breve{\Sigma}_K$ be the $2K \times 2K$ matrix

$$\breve{\Sigma}_K = \begin{pmatrix} \Sigma_K & 0_{K\times K} \\ 0_{K\times K} & \Sigma_K \end{pmatrix},$$

where $\Sigma_K$ is defined in (A.4). Part 2 of Lemma A.4 implies that $C_{NK} \to^d N(0, \breve{\Sigma}_K) \sim B_K$ as $N \to \infty$. Result (A.12) now follows from the continuous mapping theorem.

To prove (A.13), it suffices to show $\tau_{Nq_N K2} \to^p 0$ as $N \to \infty$ followed by $K \to \infty$. The same argument shows that $\tau_{N\tilde{q}_N K2} \to^p 0$ as $N \to \infty$ followed by $K \to \infty$. By Lemma A.3 in Section A.3,

$$E(\tau_{Nq_N K2}) = \sum_{k=K+1}^{\infty} E(c_{Nq_N k}^2)$$

$$= \sum_{k=K+1}^{\infty} \left\{ N^{-1/2}\left(1+\frac{m}{n}\right)\int_{z_1}\int_{z_2} D[(\min(z_1,z_2)]\psi_k(z_1)\psi_k(z_2)d\mu(z_1)d\mu(z_2) \right.$$

$$-2N^{-1/2}\left(1+\frac{m}{n}\right)\left[\int_z D(z)\psi_k(z)d\mu(z)\right]\left[\int_z F_Y(z;\mathcal{O})\psi_k(z)d\mu(z)\right]+(\Sigma_K)_{k,k}\left(2+\frac{n}{m}+\frac{m}{n}\right)\frac{\lambda}{(\lambda+1)^2}$$

$$\left. + N^{-2}\left[\int_z D(z)\psi_k(z)d\mu(z)\right]^2 \sum_{i=1}^{n}\sum_{j=1}^{n} E(U_{iq_N}U_{jq_N}) \right\}.$$

The last expression is bounded as $N \to \infty$ for every positive integer $K$, which implies that

$$\lim_{K\to\infty}\lim_{N\to\infty} E(\tau_{Nq_N K2}) = 0.$$

The result (A.13) follows from this and Markov's inequality. Q.E.D.

Proof of Theorem 3.2: Arguments like those used to prove Lemma A.2 show that $(\tau_{nmq_N}, \tau_{nm\tilde{q}_N})/N \to^p (0,0)$. Theorem 3.2 follows from this result. Q.E.D.

Proofs of Theorem 3.3: Same as the proof of Theorem 3.2 after replacing $\tau_{nm}$ with $\nu_{nm}$. Q.E.D.

Proofs of Theorem 4.1 and 4.2: These theorems follow from arguments similar to those used to prove Theorems 3.2-3.3. Q.E.D.

A.2 Auxiliary Lemmas

Define $D$ and $N$ as in the paragraph preceding Lemma A.2.

Lemma A.3: Let assumption 2 hold, and let $q_N$ and $\tilde{q}_N$ be two permutations of $\{1,...,N\}$ that are sampled independently from the uniform distribution on $\{1,2,...,N\}$. Let $i_q$ denote the position of observation $i$ ($i = 1,...,N$) in permutation $q$ of the original sample. Define

$$U_{iq_N} = (N/n)I(i_{q_N} \le n) - (N/m)I(i_{q_N} > n).$$

Define $U_{i\tilde{q}_N}$ similarly with $\tilde{q}_N$ in place of $q_N$. Then as $N \to \infty$,

(A.14) $N^{-1/2}\sum_{i=1}^{n} U_{iq_N} = O_p(1)$,

(A.15) $N^{-1}\sum_{i=1}^{n} U_{iq_N}^2 \to^p (\lambda+1)/\lambda$,

(A.16) $N^{-1}\sum_{i=n+1}^{N} U_{iq_N}^2 \to^p \lambda+1$,

(A.17) $N^{-1}\sum_{i=1}^{n} U_{iq_N} U_{i\tilde{q}_N} \to^P 0$,

(A.18) $N^{-1}\sum_{i=n+1}^{N} U_{iq_N} U_{i\tilde{q}_N} \to^P 0$,

(A.19) $E\left(\sum_{i=1}^{n} U_{iq_N}\right) = E\left(\sum_{i=n+1}^{N} U_{iq_N}\right) = 0$,

(A.20) $N^{-1}E\left(\sum_{i=1}^{n} U_{iq_N}^2\right) = 1+n/m$,

and

(A.21) $N^{-1}E\left(\sum_{i=n+1}^{N} U_{iq_N}^2\right) = 1+m/n$. ■

Proof: We begin by obtaining preliminary results that are used to prove (A.14)-(A.21). The quantity

$$n^{-1}\sum_{i=1}^{n} I(i_{q_N} \le n)$$

has a hypergeometric distribution for $n$ draws from a population of size $N$ that has $n$ "successes." Therefore,

$$En^{-1}\sum_{i=1}^{n} I(i_{q_N} \le n) = n/N \to (1+\lambda)^{-1}$$

and

$$Var\left[n^{-1}\sum_{i=1}^{n} I(i_{q_N} \le n)\right] = \frac{m^2}{N^2(N-1)} \to 0$$

as $N \to \infty$. It follows that

$$n^{-1}\sum_{i=1}^{n} I(i_{q_N} \le n) \to^p (1+\lambda)^{-1}.$$

By a similar argument,

$$m^{-1}\sum_{i=n+1}^{N} I(i_{q_N} \le n) \to^p (1+\lambda)^{-1}.$$

In addition, Theorem 1 of Lahiri, Chatterjee, and Matti (2007) implies that

$$N^{1/2}\left[N^{-1}\sum_{i=1}^{n} I(i_{q_N} \le n) - (n/N)^2\right] = O_p(1)\,.$$

Now consider the limiting behavior of

$$n^{-1}\sum_{i=1}^{n} I(i_{q_N} \le n) I(i_{\tilde{q}_N} \le n)\,.$$

tFix $\bar{i}_2 \in \{0,...,n\}$ arbitrarily. Consider the even that out of the observations indexed by $i = 1,...,n$, there are exactly $\bar{i}_2$ such that $I(i_{\tilde{q}_N} \le n) = 1$. By the hypergeometric distribution, the probability of this event is

$$\binom{n}{\bar{i}_2}\binom{m}{n-\bar{i}_2}\binom{N}{n}^{-1}.$$

In addition, because the permutations $q_N$ and $\tilde{q}_N$ are independent, $I(i_{q_N} \le n) I(i_{\tilde{q}_N} \le n)$ has the hypergeometric distribution, and so

$$E_{\bar{i}_2}\left[n^{-1}\sum_{i=1}^{n} I(i_{q_N} \le n) I(i_{\tilde{q}_N} \le n)\right] = \frac{n\bar{i}_2}{N},$$

and

$$Var_{\bar{i}_2}\left[n^{-1}\sum_{i=1}^{n} I(i_{q_N} \le n) I(i_{\tilde{q}_N} \le n)\right] = n\frac{\bar{i}_2}{N}\frac{N-\bar{i}_2}{N}\frac{m}{N-1},$$

where $E_{\bar{i}_2}$ and $Var_{\bar{i}_2}$, respectively, denote the mean and variance conditional on

$$\sum_{i=1}^{n} I(i_{\tilde{q}_N} \le n) = \bar{i}_2.$$

The unconditional mean is

$$E\left[n^{-1}\sum_{i=1}^{n} I(i_{q_N} \le n) I(i_{\tilde{q}_N} \le n)\right] = (n/N)^2 \to (1+\lambda)^{-2}\,.$$

The unconditional variance satisfies

$$Var\left[n^{-1}\sum_{i=1}^{n} I(i_{q_N} \le n) I(i_{\tilde{q}_N} \le n)\right] \to 0\,.$$

Therefore,

$$n^{-1}\sum_{i=1}^{n} I(i_{q_N} \le n) I(i_{\tilde{q}_N} \le n) \to^{p} (1+\lambda)^{-2}\,.$$

By an analogous argument,

$$m^{-1}\sum_{i=n+1}^{N} I(i_{q_N} \le n) I(i_{\tilde{q}_N} \le n) \to^p (1+\lambda)^{-2}.$$

We now use the foregoing results to prove (A.14)-(A.21). Result (A.14) now follows from

$$N^{-1/2}\sum_{i=1}^{n} U_{iq_N} = (N/n + N/m)N^{1/2}\left[N^{-1}\sum_{i=1}^{N} I(i_{q_N} \le n) - (n/N)^2\right] = O_p(1).$$

Result (A.15) follows from

$$N^{-1}\sum_{i=1}^{n} U_{iq_N}^2 = (N/n^2)\sum_{i=1}^{n} I(i_{q_N} \le n) + (N/m^2)\sum_{i=1}^{n} I(i_{q_N} > n) \to^p (\lambda+1)/\lambda.$$

A similar argument gives (A.16). Result (A.17) follows from

$$N^{-1}\sum_{i=1}^{N} U_{iq_N} U_{i\tilde{q}_N} = N^{-1}\sum_{i=1}^{n}\left[\frac{N}{n} I(i_{q_N} \le n) - \frac{N}{m} I(i_{q_N} > n)\right]\left[\frac{N}{n} I(i_{\tilde{q}_N} \le n) - \frac{N}{m} I(i_{\tilde{q}_N} > n)\right]$$

$$\to^p \left[\frac{1+\lambda}{\lambda^2} - 2\frac{1+\lambda}{\lambda^2} + \frac{1+\lambda}{\lambda^2}\right] = 0.$$

A similar argument yields (A.18).

To obtain (A.19) observe that

$$\sum_{i=1}^{n} U_{iq_N} = \frac{N}{n}\sum_{i=1}^{n} I(i_{q_N} \le n) - \frac{N}{m}\sum_{i=1}^{n} I(i_{q_N} > n).$$

This and the preliminary results imply that

$$E\left(\sum_{i=1}^{n} U_{iq_N}\right) = 0.$$

This and

$$\sum_{i=1}^{N} U_{iq_N} = 0$$

imply that

$$E\left(\sum_{i+n+1}^{N} U_{iq_N}\right) = 0,$$

which establishes (A.19).

To prove (A.20), observe that

$$N^{-1}\sum_{i=1}^{n} U_{iq_N}^2 = \frac{N}{n^2}\sum_{i=1}^{n} I(i_{q_N} \le n) + \frac{N}{m^2}\sum_{i=1}^{n} I(i_{q_N} > n).$$

This result and the preliminary results imply that

$$E\left[ N^{-1}\sum_{i=1}^{n} U_{iq_N}^2 \right] = 1 + n/m .$$

In addition,

To prove (A.21), observe that

$$N^{-1}\sum_{i=1}^{N} U_{iq_N}^2 = \frac{N}{n^2}\sum_{i=1}^{N} I(i_{q_N} \le n) + \frac{N}{m^2}\sum_{i=1}^{N} I(i_{q_N} > n)$$

$$= 2 + n/m + m/n.$$

This result and (A.20) imply (A.21). Q.E.D.

Lemma A.4: Let assumptions 1 and 2 hold, $q_N$ and $\tilde{q}_N$ be two permutations of $\{1,...,N\}$ that are sampled independently from the uniform distribution on $\{1,2,...,N\}$, $\hat{F}_{Xq_N}$ ($\hat{F}_{X\tilde{q}_N}$) be the empirical distribution function of the first $n$ observations in permutation $q_N$ ($\tilde{q}_N$), and $\hat{F}_{Yq_N}$ ($\hat{F}_{Y\tilde{q}_N}$) be the empirical distribution function of observations $n+1,...,N$. Then

(A.22) $$N^{1/2}\int [\hat{F}_X(z;\mathcal{O}) - \hat{F}_Y(z;\mathcal{O})]\{\psi_k(z)\}_{k=1}^{K} d\mu(z) \to^d N(\Xi, \Sigma_K)$$

and

(A.23) $$N^{1/2}\begin{bmatrix} \int [\hat{F}_{Xq_N}(z;\mathcal{O}) - \hat{F}_{Yq_N}(z;\mathcal{O})]\{\psi_k(z)\}_{k=1}^{K} d\mu(z) \\ \int [\hat{F}_{X\tilde{q}_N}(z;\mathcal{O}) - \hat{F}_{Y\tilde{q}_N}(z;\mathcal{O})]\{\psi_k(z)\}_{k=1}^{K} d\mu(z) \end{bmatrix} \to^d N\left( \boldsymbol{0}_{2K\times 1}, \begin{bmatrix} \Sigma_K & \boldsymbol{0}_{K\times K} \\ \boldsymbol{0}_{K\times K} & \Sigma_K \end{bmatrix} \right),$$

where

$$\Xi = \int D(z)\{\psi_k(z)\}_{k=1}^{K} d\mu(z)$$

and $\Sigma_K$ is the $K \times K$ matrix defined in (A.4). ■

Proof: Let $\{W_i : i = 1,...,N\}$ denote the combined sample of observations of $X$ and $Y$.

Proof of (A.23): Let $i_q$ denote the position of observation $i$ ($i = 1,...,N$) in permutation $q$ of the original sample. Then for any permutation $q$,

(A.24) $$N^{1/2}[\hat{F}_{Xq}(z;\mathcal{O}) - \hat{F}_{Yq}(z;\mathcal{O})] = N^{-1/2}\sum_{i=1}^{N} U_{iq} I[W_i(t) \le z(t) \;\; \forall\, t \in \mathcal{O}],$$

where

$$U_{iq} = (N/n)I(i_q \le n) - (N/m)I(i_q > n) .$$

Step 1: We show that

(A.25) $$N^{-1/2}\sum_{i=1}^{N}\begin{pmatrix} U_{iq_N} \\ U_{i\tilde{q}_N} \end{pmatrix}\left\{\int I[W_i(t)\le z(t)\ \ \forall\, t\in\mathcal{O}]\psi_k(z)d\mu(z)-\mu_{iNk}\right\}_{k=1}^{K}\to^d N\left(\boldsymbol{0}_{2K\times 1},\begin{bmatrix}\Sigma_K & \boldsymbol{0}_{K\times K}\\ \boldsymbol{0}_{K\times K} & \Sigma_K\end{bmatrix}\right),$$

where

$$\mu_{iNk}=I(i\le n)\int F_{nX}(z;\mathcal{O})\psi_k(z)d\mu(z)+I(i>n)\int F_Y(z;\mathcal{O})\psi_k(z)d\mu(z)\,.$$

Let $\alpha,\beta\in\mathbb{R}$ and $\gamma\in\mathbb{R}^K$ be arbitrary constants. By the Cramér-Wold device, it suffices to show that

(A.26) $$N^{-1/2}\sum_{i=1}^{N}\Upsilon_i\to^d N(0,\sigma^2)\,,$$

where

$$\Upsilon_i=\sum_{k=1}^{K}\gamma_k(\alpha U_{iq_N}+\beta U_{i\tilde{q}_N})\left\{\int I[W_i(t)\le z(t)\ \ \forall\, t\in\mathcal{O}]\psi_k(z)d\mu(z)-\mu_{iNk}\right\}$$

and

$$\sigma^2=(\alpha^2+\beta^2)\sum_{k,\tilde{k}=1}^{K}\gamma_k\tilde{\gamma}_{\tilde{k}}\Sigma_{K,k\tilde{k}}\,.$$

To establish (A.26), observe that conditional on $(q_N,\tilde{q}_N)$, $\{\Upsilon_i\}_{i=1}^N$ is a sequence of independent mean-zero random variables with variances

$$\sigma_{iN}^2=\sum_{k,\tilde{k}=1}^{K}\gamma_k\gamma_{\tilde{k}}(\alpha U_{iq_N}+\beta U_{i\tilde{q}_N})^2\Big\{I(i\le n)\int_z\int_{\tilde{z}}\{F_{nX}[\min(z,\tilde{z});\mathcal{O}]-F_{nX}(z;\mathcal{O})F_{nX}(\tilde{z};\mathcal{O})\}\psi_k(z)\psi_{\tilde{k}}(\tilde{z})d\mu(z)d\mu(\tilde{z})$$

$$+I(i>n)\int_z\int_{\tilde{z}}\{F_Y[\min(z,\tilde{z});\mathcal{O}]-F_Y(z;\mathcal{O})F_Y(\tilde{z};\mathcal{O})\}\psi_k(z)y_{\tilde{k}}(\tilde{z})d\mu(z)d\mu(\tilde{z})\Big\}.$$

By Lemma A.3,

(A.27) $$\sigma_N^2\equiv N^{-1}\sum_{i=1}^{N}\sigma_{iN}^2\to\sigma^2>0$$

with probability 1 relative to the distribution of $(q_N,\tilde{q}_N)$. Moreover, for any sufficiently small $\delta>0$ and as $N\to\infty$,

$$E(|\Upsilon_i|^{2+\delta}|\,q_N,\tilde{q}_N)=K^{2+\delta}\max_{k\le K}|\gamma_k|^{2+\delta}[\max(|\alpha|,|\beta|)]^{2+\delta}\left[\frac{N}{\min(n,m)}\right]^{2+\delta}\max_{\tilde{k}\le K}\left[\int|\psi_{\tilde{k}}(z)|\,d\mu(z)\right]^{2+\delta}$$

(9.28) $$\to K^{2+\delta}\max_{k\le K}|\gamma_k|^{2+\delta}[\max(|\alpha|,|\beta|)]^{2+\delta}\left[\frac{\lambda+1}{\min(1,\lambda)}\right]^{2+\delta}\max_{\tilde{k}\le K}\left[\int|\psi_{\tilde{k}}(z)|\,d\mu(z)\right]^{2+\delta}<\infty.$$

Result (A.26) and, therefore, (A.25), now follows from (A.27), (A.28), and a triangular array central limit theorem (Serfling 1980, p. 30).

Step 2: For any $k = 1,...,K$

(A.29) $$N^{-1/2}\sum_{i=1}^{N}(U_{iq_N},U_{i\tilde{q}_N})\mu_{iNk} = N^{-1/2}\sum_{i=1}^{n}(U_{iq_N},U_{i\tilde{q}_N})\int[F_{nX}(z;\mathcal{O}) - F_Y(z;\mathcal{O})]\psi_k(z)d\mu(z),$$

where we have used

$$\sum_{i=1}^{N}U_{iq_N} = \sum_{i=1}^{N}U_{i\tilde{q}_N} = 0 .$$

Lemma A.3 implies that

$$N^{-1/2}\sum_{i=1}^{n}(U_{iq_N},U_{i\tilde{q}_N}) = O_p(1) .$$

Therefore, the right-hand side of (A.29) is $O_p(1)$. Combining this result, (A.24), and (A.25) yields (A.23). Proof of (A.22): We have

(A.30) $$N^{1/2}[\hat{F}_X(z;\mathcal{O}) - \hat{F}_Y(z;\mathcal{O})] = N^{-1/2}\sum_{i=1}^{N}U_i I[W_i(t) \le z(t) \ \ \forall\, t \in \mathcal{O}],$$

where

$$U_i = (N/n)I(i \le n) - (N/m)I(i > n) .$$

By an argument similar to that used in step 1 of the proof of (A.23), we can show that for any $K \ge 1$

(A.31) $$N^{-1/2}\sum_{i=1}^{N}U_i\left\{\int I[W_i(t) \le z(t) \ \ \forall\, t \in \mathcal{O}]\psi_k(z)d\mu(z) - \mu_{iNk}\right\}_{k=1}^{K} \to^d N(\boldsymbol{0}_{K\times 1},\Sigma) .$$

Also, by an argument similar to step 2 of the proof of (A.23), we can show that

(A.32) $$N^{-1/2}\sum_{i=1}^{N}U_i\mu_{iNk} = N^{1/2}\int D(z)\psi_k(z)d\mu(z) .$$

Result (A.22) follows from (A.30)-(A.32). Q.E.D.

A.3. Asymptotic Distributions under Local Alternatives

We first consider the asymptotic local power of the $\tau_{nm}$ permutation test when $X(t)$ and $Y(t)$ are observed at a the finite set of points $(t_1,...,t_J)$. Let $\tilde{\Upsilon}(\zeta)$ be the Gaussian process indexed by $\zeta \in \mathbb{R}^J$ defined in the paragraph preceding Lemma A.2. Define a sequence of local alternatives by

$$F_{nX}(\zeta) = F_Y(\zeta) + (n+m)^{-1/2}\tilde{D}(\zeta)$$

for every $\zeta \in \mathbb{R}^J$ and some function $\tilde{D}$ such that $\int \tilde{D}(\zeta)^2 d\mu_J < \infty$. $F_X$ is now indexed by the sample size $n$ because, under a sequence of local alternatives, $F_X$ changes as $n$ increases. $F_Y$ can also be indexed by $m$. We do not index $F_Y$ this way because doing so adds complexity to the notation without changing the result. Define $r(1-\alpha)$ as the $1-\alpha$ quantile of the distribution of the random variable $\int [\tilde{\Upsilon}(\zeta)]^2 d\mu_J$.

The following theorem gives the asymptotic power of the $\tau_{nm}$ permutation test against sequences of local alternatives when $X(t)$ and $Y(t)$ are observed at a finite set of points.

Theorem A.1: Let assumptions 1 and 2 hold. Then,

$$P\left\{\int [\tilde{\Upsilon}(\zeta) + \tilde{D}(\zeta)]^2 d\mu_J > r(1-\alpha)\right\} \le \liminf_{n\to\infty} P[\tau_{nm} > t^*_{nm}(1-\alpha)]$$

$$\le \limsup_{n\to\infty} P[\tau_{nm} > t^*_{nm}(1-\alpha)]$$

$$\le \lim_{\delta\to 0^+} P\left\{\int [\tilde{\Upsilon}(\zeta) + \tilde{D}(\zeta)]^2 d\mu_J > r(1-\alpha) - \delta\right\}. \blacksquare$$

Proof of Theorem A.1: This theorem follows from Proposition A.1 and Lemma A.2. Q.E.D.

It follows from Theorem A.1 that the $\alpha$-level permutation test based on $\tau_{nm}$ has asymptotic local power exceeding $\alpha$ whenever $\int [\tilde{D}(\zeta)]^2 d\mu_J > 0$.

We now consider the asymptotic local power of the $\tau_{nm}$ test when $X(t)$ and $Y(t)$ are observed in continuous time. $\Upsilon(z)$ is the Gaussian process indexed by $z \in L_2(\mu)$ defined in the paragraph preceding Lemma A.2. Define a sequence of local alternatives by

$$F_{nX}(z) = F_Y(z) + (n+m)^{-1/2} D(z)$$

for every $z \in L_2(\mu)$ and some function $D$ such that $\int D(z)^2 d\mu < \infty$. As in the discrete case, $F_X$ is indexed by $n$ because, under a sequence of local alternatives, $F_X$ changes as $n$ increases. Define $r(1-\alpha)$ as the $1-\alpha$ quantile of the distribution of the random variable $\int \Upsilon(z)^2 d\mu$.

The following theorem gives the asymptotic power of the permutation test against sequences of local alternatives when $X(t)$ and $Y(t)$ are observed in continuous time.

Theorem A.2: Let assumptions 1 and 2 hold. Then,

$$P\left\{\int [\Upsilon(z)+D(z)]^2 d\mu > r(1-\alpha)\right\} \le \liminf_{n\to\infty} P[\tau_{nm} > t^*_{nm}(1-\alpha)]$$

$$\le \limsup_{n\to\infty} P[\tau_{nm} > t^*_{nm}(1-\alpha)]$$

$$\le \lim_{\delta\to 0^+} P\left\{\int [\Upsilon(z)+D(z)]^2 d\mu > r(1-\alpha)-\delta\right\}. \blacksquare$$

Proof of Theorem A.2: Like Theorem A.1, Theorem A.2 follows from Proposition A.1 and Lemma A.2. Q.E.D.

Now consider the $\nu_{nm}$ test when $X(t)$ and $Y(t)$ are observed in continuous time. Define a sequence of local alternatives by

(A.33) $X(t) = Y(t) + (n+m)^{-1/2} D(t)$.

where $D$ is a non-stochastic function and

(A.34) $\int_0^T D(t)^2 dt < \infty$.

Under (A.33) and (A.34), arguments like those used to prove Theorems A.1 and A.2 apply to $\nu_{nm}$ after replacing the processes $n^{1/2}[\hat{F}_X(t) - F_X(t)]$ and $m^{1/2}[\hat{F}_Y(t) - F_Y(t)]$ with $n^{1/2}[\hat{E}X(t) - EX(t)]$ and $m^{1/2}[\hat{E}Y(t) - EY(t)]$, respectively. Therefore,

Theorem A.3: Let assumptions 1-3 hold. Let $\tilde{\Upsilon}(t)\,(0 \le t \le T)$ be a mean-zero Gaussian process with the same covariance function as $X(t)$ and $Y(t)$. Let $r(1-\alpha)$ denote the $1-\alpha$ quantile of $\int_0^T \tilde{\Upsilon}(t)^2 dt$. Then

$$P\left\{\int_0^T [\tilde{\Upsilon}(t)+D(t)]^2 dt > r(1-\alpha)\right\} \le \liminf_{n\to\infty} P[\nu_{nm} > \nu^*_{nm}(1-\alpha)]$$

$$\le \limsup_{n\to\infty} P[\nu_{nm} > \nu^*_{nm}(1-\alpha)]$$

$$\le \lim_{\delta\to 0^+} P\left\{\int_0^T [\tilde{\Upsilon}(t)+D(t)]^2 dt > r(1-\alpha)-\delta\right\}. \blacksquare$$

A similar result holds when $X(t)$ and $Y(t)$ are observed in discrete time.

A.4. Sensitivity of the $\tau_n$, $\eta_n$, SR, and HT tests to various parameters.

Table A1 presents the results of Monte Carlo experiments that illustrate the sensitivity of the power of the $\tau_n$ and SR tests to the frequency of the observations in the discrete case. Table A2 illustrates the sensitivity of the power of the HT test to its tuning parameters. Figures A1 and A2 illustrate the sensitivity of the power of the $\eta_n$ test to $K$. Because of the long computing times required to simulate a month of observations at several observation frequencies, Table A1 is based on one simulated day of observations. For brevity, Table A1 is based only on designs 4 and 6 of Section 7.

## REFERENCES

Andrews, D.W.K. and P. Guggenberger (2010). Asymptotic size and a problem with subsampling with the m out of n bootstrap. *Econometric Theory*, 26, 426–468.

Bertanha, M. and E. Chung (2021). Permutation tests at nonparametric rates. arXiv preprint arXiv:2102.13638.

Bugni, F.A., P. Hall, J.L. Horowitz, and G.R. Neumann (2009). Goodness-of-fit tests for functional data. *The Econometrics Journal*, 12. S1-S18.

Chung, E. and J. P. Romano (2013). Exact and asymptotically robust permutation tests. *Annals of Statistics*, 41, 484-507.

Chung, E. and J.P. Romano (2016). Multivariate and multiple permutation tests. *Journal of Econometrics*, 193, 76-91.

Cuesta-Albertos, J.A., R. Fraiman, and T. Ransford (2006). Random projections and goodness of fit tests in infinite dimensional spaces. *Bulletin of the Brazilian Mathematical Society*, 37, 1-25.

Cuesta-Albertos, E. del Barrio, R. Fraiman, and C. Matrán (2007). The random projection method in goodness of fit for functional data. *Computational Statistics and Data Analysis*, 51, 4814-4831.

Fogarty, C. B. (2020). Studentized sensitivity analysis for the sample average treatment effect in paired observational studies. *Journal of the American Statistical Association*, 115, 1518-1530.

Fogarty, C. B. (2021). Prepivoted permutation tests. arXiv preprint arXiv:2102.04423.1.

Hall, P. and M. Hosseini-Nasab (2006). On properties of functional principal components analysis. *Journal of the Royal Statistical Society*, Series B., 68, 109-126.

Hall, P. and N. Tajvidi (2002). Permutation tests for equality of distributions in high-dimensional settings. *Biometrika*, 89, 359-374.

Harding, M. and C. Lamarche (2016). Empowering consumers through smart technology: experimental evidence on the consequences of time-of-use electricity pricing. *Journal of Policy Analysis and* Management, 35, 906-931.

Henze, N. (1988) A multivariate two-sample test based on the number of nearest neighbor type coincidences. *Annals of Statistics*, 16, 772-783.

Hoeffding, W. (1952). The large-sample power of tests based on permutations of observations. *Annals of Mathematical Statistics*, 3, 169-192.

Jank, W. and G. Shmueli (2006). Functional data analysis in electronic commerce research. *Statistical* Science, 21, 155-166.

Kim, M.S. and S. Wang (2006). Sizes of two bootstrap-based nonparametric specification tests for the drift function in continuous time models. *Computational Statistics and Data Analysis*, 50, 1793-1806.

Lahiri, S.N., A. Chatterjee, and T. Matti (2007). Normal approximation to the hypergeometric distribution in nonstandard cases and a sub-Gaussian Berry-Esséen theorem. *Journal of Statistical Planning and Inference*, 137, 3570-3590.

Lehmann, E.L. and J.P. Romano (2005). *Testing Statistical Hypotheses*, 3rd edition, New York: Springer.

Ramsay, J.O. and B.J. Silverman (2002). *Applied Functional Data Analysis: Methods and Case Studies.* New York: Springer.

Ramsay, J.O. and B.W. Silverman (2005). *Functional Data Analysis*. New York: Springer.

Romano, J.P. (1989). Bootstrap and randomization tests of some nonparametric hypotheses. *Annals of Statistics*, 17, 141-159.

Schilling, M. (1986). Multivariate two-sample tests based on nearest neighbors. *Journal of the American Statistical Association*, 81, 799-806.

Serfling, R.J. (1980). *Approximation Theorems in Mathematical Statistics*. New York: Wiley.

Székely, G. and M. Rizzo (2004). Testing for equal distributions in high-dimension. Working paper, Bowling Green State University and Ohio University.

Yao, F., H.-G. Müller, and J.-L. Wang (2005). Functional linear regression analysis for longitudinal data. *Annals of Statistics*, 33, 2873-2903.

### TABLE 1: DISTRIBUTION OF CUSTOMERS AMONG GROUPS

| | Control | Treatment 1 | Treatment 2 | Treatment 3 | Treatment 4 | Total |
|---|---|---|---|---|---|---|
| Number of Customers | 524 | 236 | 227 | 251 | 254 | 1492 |
| Percentage of Customers | 35.1 | 15.8 | 15.2 | 16.8 | 17.0 | 100 |

### TABLE 2: *P*-VALUES IN % OF THE TESTS WITH THE CBT DATA[a]

| Test | $(\alpha_\tau, \alpha_V)$ | June | July | Aug. | Sept. | Oct. | Nov. | Dec. |
|---|---|---|---|---|---|---|---|---|
| $\eta_n$ | (0.04,0.01) | 14.4 | 6.4 | 1.9 | 10.9 | 29.7 | 51.7 | 100 |
| | (0.03,0.02) | 15.3 | 8.6 | 2.6 | 14.5 | 39.6 | 68.9 | 100 |
| | (0.025.0.025) | 12.3 | 10.3 | 3.1 | 17.4 | 47.5 | 82.7 | 100 |
| | (0.02,0.03) | 10.2 | 12.8 | 3.8 | 21.8 | 59.3 | 100 | 100 |
| $\tau_n$ | N.A | 11.6 | 5.2 | 1.6 | 8.7 | 23.8 | 41.4 | 84.4 |
| SR | N.A. | 11.8 | 18.6 | 30.4 | 16.8 | 92.2 | 88.4 | 93.6 |

a. As is explained in Section 6, the $p$-values in this table are the values of $\alpha$ in $(\alpha_V, \alpha_\tau)\alpha / 0.05$ at which $\eta_n$ switches from 0 to 1. The powers of $\eta_n$ and $\tau_n$ are for $K = 25$ .

## TABLE 3: EMPIRICAL REJECTION PROBABILITIES IN % IN THE MONTE CARLO EXPERIMENTS

| Test | $(\alpha_\tau, \alpha_V)$ | Design 1 | Design 2 | Design 3 | Design 4 | Design 5 | Design 6 | Design 7 | Design 8 | Design 9 | Design 10 |
|---|---|---|---|---|---|---|---|---|---|---|---|
| $\eta_n$ | (0.04,0.01) | 4.3 | 42.8 | 55.7 | 83.2 | 87.1 | 64.4 | 95.8 | 64.3 | 96.7 | 88.9 |
| | (0.03, 0.02) | 4.4 | 51.7 | 62.1 | 83.3 | 83.6 | 59.7 | 93.3 | 60.2 | 94.6 | 84.0 |
| | (0.025, 0.025) | 4.3 | 52.7 | 63.1 | 82.8 | 82.7 | 55.3 | 91.8 | 56.2 | 93.4 | 81.9 |
| | (0.02, 0.03) | 4.4 | 55.2 | 64.7 | 82.1 | 80.2 | 50.9 | 90.5 | 52.4 | 91.9 | 77.8 |
| $\tau_n$ | N.A. | 4.9 | 18.8 | 41.0 | 78.4 | 84.9 | 66.8 | 95.6 | 66.7 | 97.6 | 89.2 |
| SR | N.A. | 5.1 | 61.8 | 61.7 | 83.9 | 56.1 | 20.7 | 20.7 | 5.6 | 19.2 | 5.3 |

**TABLE A1: SENSITIVITY OF THE POWERS IN % OF THE $\tau_{nm}$ AND SR TESTS TO THE FREQUENCY OF DISCRETE OBSERVATIONS[a]**

| Design | Frequency | $\tau_{nm}$ | SR |
|---|---|---|---|
| 4 | 5 min | 80.7 | 23.4 |
| | 10 min | 61.2 | 18.3 |
| | 30 min | 30.4 | 9.8 |
| 6 | 5 min | 80.4 | 24.4 |
| | 10 min | 60.6 | 18.6 |
| | 30 min | 30.2 | 10.8 |

a. The designs are described in Section 7. The powers of the $\tau_{nm}$ test are with $K = 25$.

**TABLE A2: SENSITIVITY OF THE POWER IN % OF THE HT TEST TO ITS TUNING PARAMETERS[a]**

| | | Design | | | | | | | | | |
|---|---|---|---|---|---|---|---|---|---|---|---|
| $\gamma$ | Weight | 1 | 2 | 3 | 4 | 5 | 6 | 7 | 8 | 9 | 10 |
| 1 | (i) | 5.4 | 11.9 | 11.4 | 100 | 26.6 | 100 | 100 | 19.4 | 100 | 17.7 |
| 1 | (ii) | 5.6 | 11.8 | 11.0 | 100 | 19.8 | 100 | 100 | 14.0 | 100 | 12.4 |
| 1 | (iii) | 5.3 | 11.3 | 11.5 | 100 | 39.1 | 100 | 100 | 29.9 | 100 | 30.4 |
| 1.5 | (i) | 5.5 | 12.5 | 12.3 | 100 | 25.3 | 100 | 100 | 17.9 | 100 | 16.6 |
| 1.5 | (ii) | 6.0 | 12.3 | 12.1 | 100 | 17.6 | 100 | 100 | 11.1 | 100 | 10.6 |
| 1.5 | (iii) | 5.7 | 11.7 | 11.6 | 100 | 42.3 | 100 | 100 | 32.5 | 100 | 32.0 |
| 2 | (i) | 5.5 | 12.7 | 12.7 | 100 | 25.1 | 100 | 100 | 17.3 | 100 | 15.8 |
| 2 | (ii) | 5.5 | 12.6 | 12.7 | 100 | 17.4 | 100 | 100 | 10.9 | 100 | 10.0 |
| 2 | (iii) | 5.6 | 11.9 | 11.7 | 100 | 43.9 | 100 | 100 | 33.5 | 100 | 33.4 |
| 2.5 | (i) | 5.3 | 13.0 | 13.1 | 100 | 24.7 | 100 | 100 | 16.4 | 100 | 14.8 |
| 2.5 | (ii) | 5.3 | 13.2 | 13.5 | 100 | 16.9 | 100 | 100 | 10.0 | 100 | 8.8 |
| 2.5 | (iii) | 5.6 | 12.3 | 12.2 | 100 | 45.9 | 100 | 100 | 34.1 | 100 | 34.8 |

a. The tuning parameters $\gamma$, weights, and labels of the weights are defined in Hall and Tajvidi (2002).

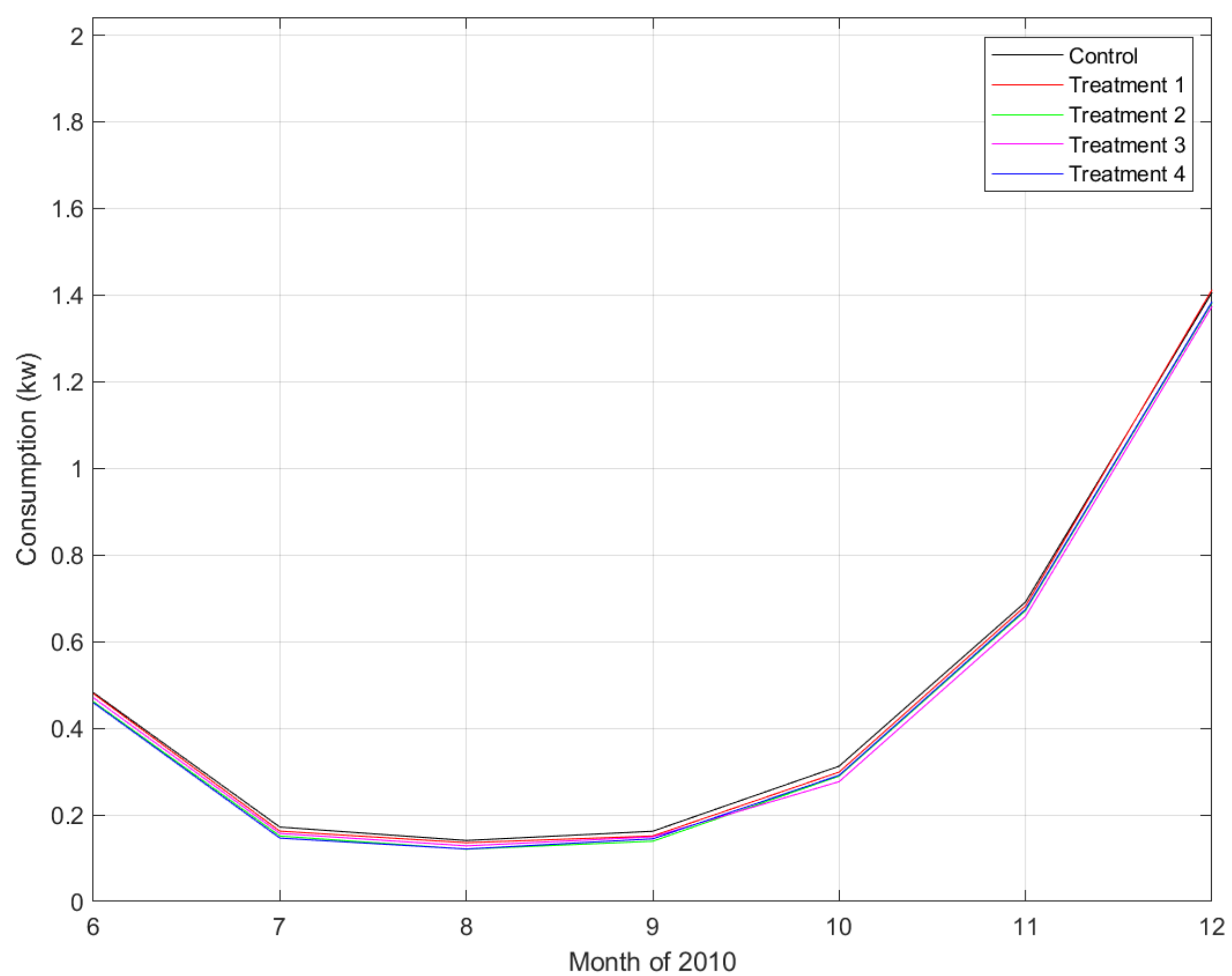

Figure 1: Average half-hour consumption

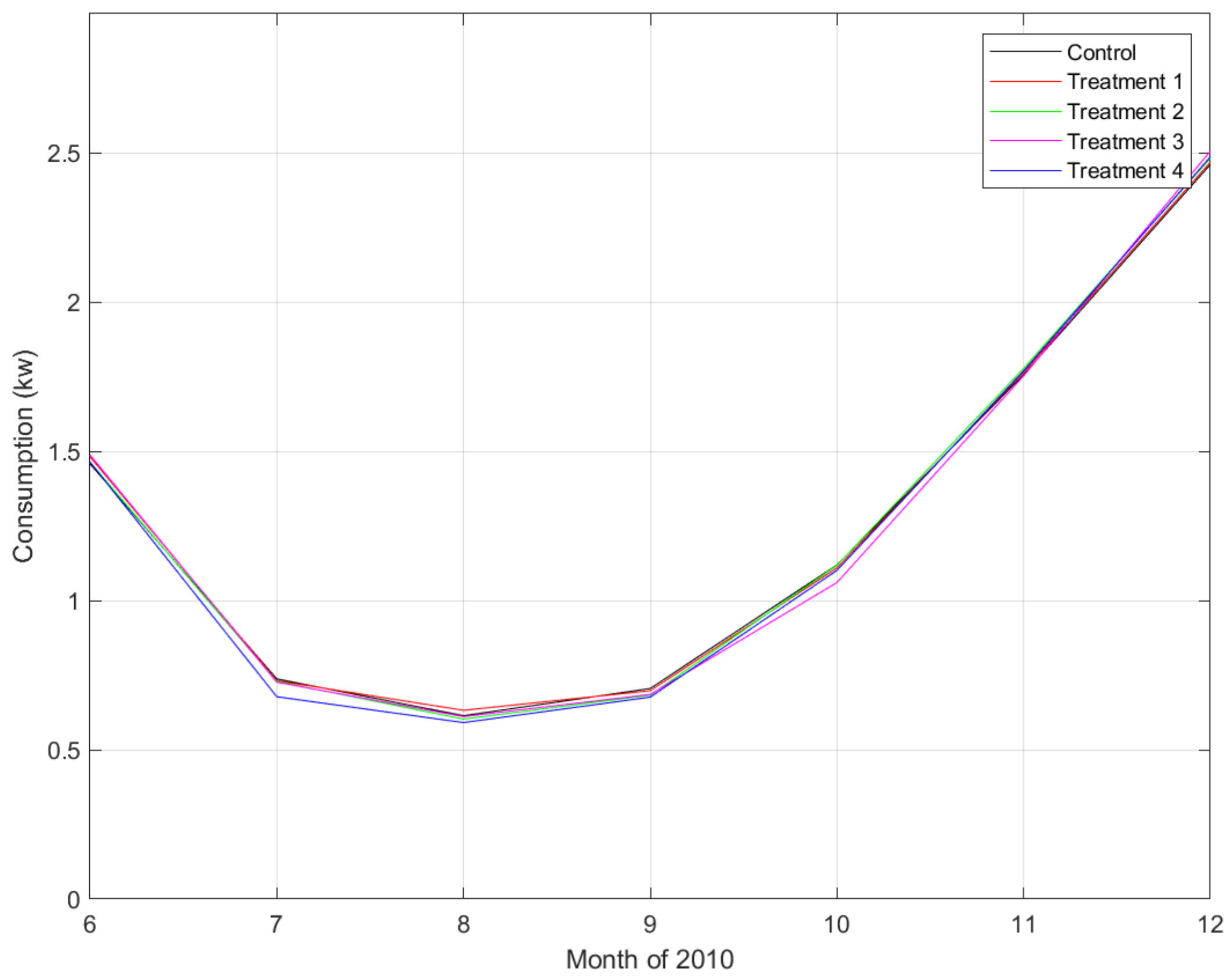

Figure 2: Sample standard deviation of half-hour consumption

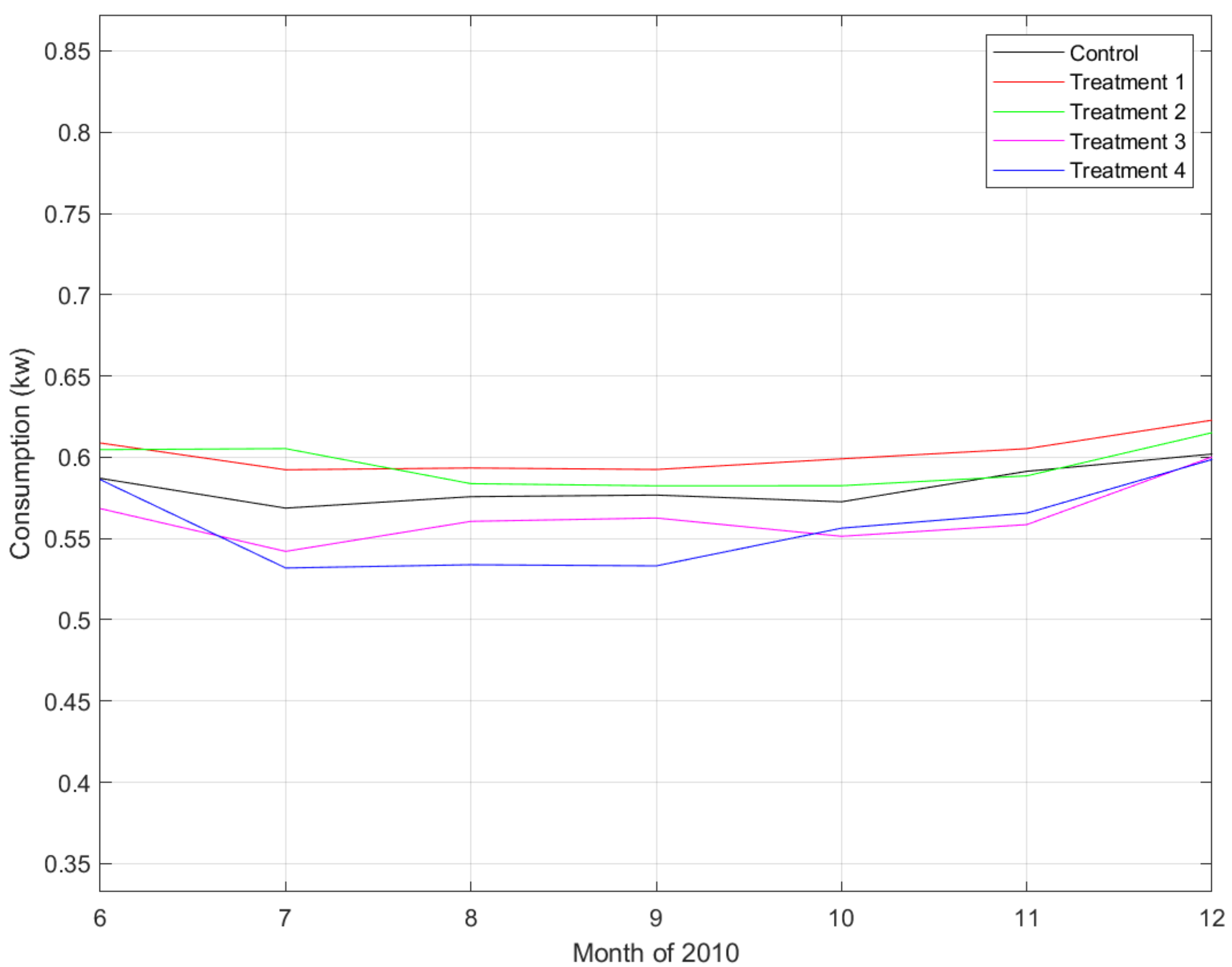

Figure 3: Sample correlation of consumption between consecutive half hours.

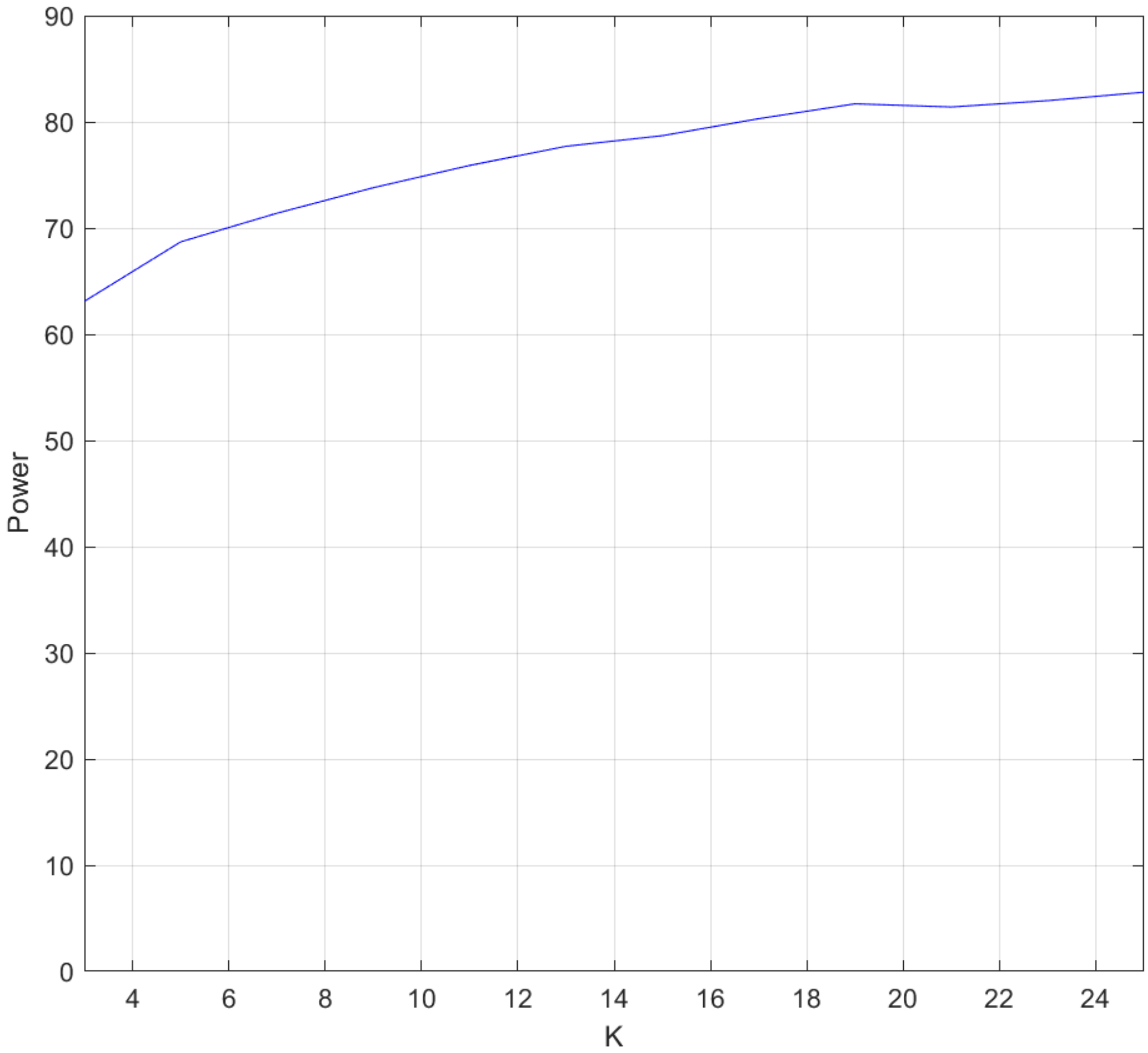

Figure A1: Power of the $\eta_n$ test in design 4 with $(\alpha_\tau, \alpha_\nu) = (0.25, 0.25)$.

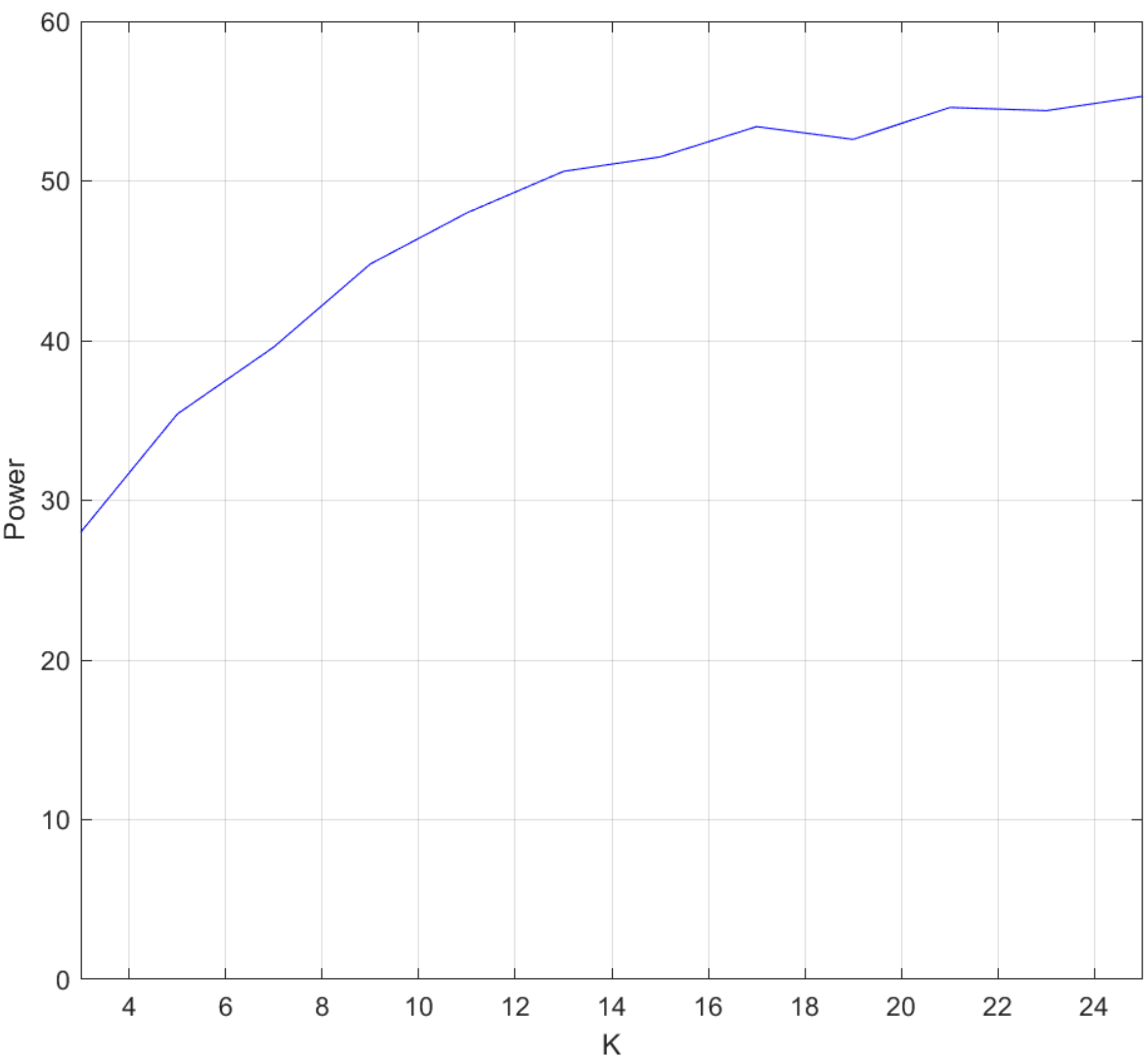

Figure A2: Power of the $\eta_n$ test in design 6 with $(\alpha_\tau, \alpha_\nu) = (0.25, 0.25)$